\documentclass[aps,prx,twocolumn,superscriptaddress,longbibliography]{revtex4-1}

\usepackage[english]{babel} 
\usepackage[latin1]{inputenc}
\usepackage[usenames,dvipsnames]{color}
\usepackage{graphicx}
\usepackage{amsmath,amssymb,amsfonts}
\usepackage[colorlinks=true,linkcolor=blue,urlcolor=blue,citecolor=blue]{hyperref}
\usepackage[percent]{overpic}

\newcommand {\mc} {\mathcal}
\newcommand {\lan} {\left \langle}
\newcommand {\ran} {\right \rangle}

\newcommand {\sign} {\mathrm{sign}}

\begin{document}

\title{Hong-Ou-Mandel heat noise in the quantum Hall regime}

\author{Flavio Ronetti}
\email{ronetti@fisica.unige.it}
\affiliation{Dipartimento di Fisica, Universit\`a di Genova, Via Dodecaneso 33, 16146, Genova, Italy}
\affiliation{CNR-SPIN, Via Dodecaneso 33, 16146, Genova, Italy}
\affiliation{Aix Marseille Univ, Universit\'e de Toulon, CNRS, CPT, Marseille, France}
\author{Luca Vannucci}
\affiliation{Dipartimento di Fisica, Universit\`a di Genova, Via Dodecaneso 33, 16146, Genova, Italy}
\affiliation{CNR-SPIN, Via Dodecaneso 33, 16146, Genova, Italy}
\affiliation{CAMD, Department of Physics, Technical University of Denmark, 2800 Kgs. 
	Lyngby, Denmark}
\author{Dario Ferraro}
\affiliation{Dipartimento di Fisica, Universit\`a di Genova, Via Dodecaneso 33, 16146, Genova, Italy}
\affiliation{Istituto Italiano di Tecnologia, Graphene Labs, Via Morego 30, I-16163 Genova, Italy}
\author{Thibaut Jonckheere}
\affiliation{Aix Marseille Univ, Universit\'e de Toulon, CNRS, CPT, Marseille, France}
\author{J\'er\^ome Rech}
\affiliation{Aix Marseille Univ, Universit\'e de Toulon, CNRS, CPT, Marseille, France}
\author{Thierry Martin}
\affiliation{Aix Marseille Univ, Universit\'e de Toulon, CNRS, CPT, Marseille, France}
\author{Maura Sassetti}
\affiliation{Dipartimento di Fisica, Universit\`a di Genova, Via Dodecaneso 33, 16146, Genova, Italy}
\affiliation{CNR-SPIN, Via Dodecaneso 33, 16146, Genova, Italy}

\begin{abstract}
We investigate heat current fluctuations induced by a periodic train of Lorentzian-shaped pulses, carrying an integer number of electronic charges, in a Hong-Ou-Mandel interferometer implemented in a quantum Hall bar in the Laughlin sequence. We demonstrate that the noise in this collisional experiment cannot be reproduced in a setup with a single drive, in contrast to what is observed in the charge noise case. Nevertheless, the simultaneous collision of two identical levitons always leads to a total suppression even for the Hong-Ou-Mandel heat noise at all filling factors, despite the presence of emergent anyonic quasi-particle excitations in the fractional regime. Interestingly, the strong correlations characterizing the fractional phase are responsible for a remarkable oscillating pattern in the HOM heat noise, which is completely absent in the integer case. These oscillations can be related to the recently predicted crystallization of levitons in the fractional quantum Hall regime.
\end{abstract}

	\maketitle

\section{Introduction}
The recent progress in generating and controlling coherent few-particle excitations in quantum conductors opened the way to a new research field, known as electron quantum optics (EQO) \cite{Bocquillon14,Grenier11b}. The main purpose of EQO is to reproduce conventional optics experiments using electronic wave-packets propagating in condensed matter systems instead of photons travelling along wave-guides.

In this context, a remarkable effort has been put forth by the condensed matter community to implement on-demand sources of electronic wave-packets in mesoscopic systems. After seminal theoretical works and groundbreaking experimental results, two main methods to realize single-electron sources assumed a prominent role in the field of EQO \cite{dubois13, grenier13, Misiorny18,Glattli16_pss,Bauerle18}. The first injection protocol relies on the periodic driving of the discrete energy spectrum of a quantum dot, which plays the role of a mesoscopic capacitor \cite{buttiker93_JPCM,buttiker93_PLA,Moskalets13_emitter}. In this way, it is possible to achieve the periodic injection of an electron and a hole along the ballistic channels of a system coupled to this mesoscopic capacitor through a quantum point contact (QPC) \cite{Feve07,Bocquillon12,Parmentier12,Ferraro15}. 

A second major step has been the recent realization of an on-demand source of electron through the application of a time-dependent voltage to a quantum conductor \cite{Glattli16_pss,Glattli16_physE,Ferraro18,Misiorny18,Moskalets16_PRL,Safi14,Dolcini17}. The main challenge to face, in this case, has been that an ac voltage would generally excite unwanted neutral electron-hole pairs, thus spoiling at its heart the idea of a single-electron source. The turning point to overcome this issue was the theoretical prediction by Levitov and co-workers that a periodic train of quantized Lorentzian-shaped pulses, carrying an integer number of particles per period, is able to inject minimal single-electron excitations devoid of any additional electron-hole pair, then termed \textit{levitons} \cite{levitov96,ivanov97,keeling06}. Indeed, this kind of single-electron source is simple to realize and operate, since it relies on usual electronic components, and potentially provides a high level of miniaturization and scalability. For their fascinating properties \cite{Moskalets15:corr}, levitons have been proposed as flying qubits \cite{Glattli18} and as source of entanglement \cite{Dasenbrook15,Dasenbrook16_NJP,Dasenbrook16_PRL,Ferraro18b} with appealing applications for quantum information processing. Moreover, quantum tomography protocols able to reconstruct their single-electron wave-functions have been proposed \cite{Grenier11,Ferraro13,Ferraro14} and experimentally realized \cite{Jullien14}. 

While the implementation of single-electron sources has not been a trivial task, the condensed matter analogues of other quantum optics experimental components can be found in a more natural way. The wave-guides for photons can be replaced by the ballistic edge channels of mesoscopic devices, such as quantum Hall systems. Moreover, the role of electronic beam splitter, which should mimic the half-silvered mirror of conventional optics, can be played by a QPC, where electrons are reflected or transmitted with a tunable probability, which is typically assumed as energy independent. By combining these elements with the single-electron sources previously described, interferometric setups, originally conceived for optics experiments, can be implemented also in the condensed matter realm \cite{Olkhovskaya08,Rossello15}. One famous example is the Hanbury-Brown-Twiss (HBT) interferometer \cite{HBT1}, where a stream of electronic wave-packets is excited along ballistic channels and partitioned against a QPC \cite{Bocquillon12}. The shot noise signal, generated due to the granular nature of electrons \cite{Martin_Houches,Moskalets17}, was employed to probe the single-electron nature of levitons in a non-interacting two-dimensional electron gas \cite{Glattli16_physE,dubois13-nature}. Its extension to the fractional quantum Hall regime was considered in Ref. \cite{Rech16}, where it was shown that levitons are minimal excitations also in strongly correlated edge channels.

A fundamental achievement of EQO has been the implementation of the Hong-Ou-Mandel (HOM) interferometer \cite{Hong87}, where electrons impinge on the opposite side of a QPC with a tunable delay \cite{Bocquillon13,Glattli16_pss,dubois13-nature}. By performing this kind of collisional experiments, it is possible to gather information about the forms of the impinging electronic wave-packets and to measure their degree of indistinguishability \cite{Jonckheere12,Ferraro18,Ferraro15}. For instance, when two indistinguishable and coherent electronic states collide simultaneously (zero time delay) at the QPC, charge current fluctuations are known to vanish at zero temperature, thus showing the so called Pauli dip \cite{Ferraro14b,Glattli16_pss,dubois13-nature}. This dip can be interpreted in terms of anti-bunching effects related to the Fermi statistics of electrons. HOM experiments can thus be employed to test whether decoherence and dephasing, induced by electron-electron interactions, reduce the degree of indistinguishability of colliding electrons \cite{Wahl14,Ferraro14,Freulon15,Marguerite16,Cabart18}.

As discussed above, the main driving force behind EQO has been to properly revise quantum optics experiments focusing on charge transport properties of single-electron excitations. Nevertheless, some recent groundbreaking experiments has spurred the investigation also in the direction of heat transport at the nanoscale \cite{Giazotto06,Pekola15,Esposito09,Campisi11,Benenti16,Li12}. In this context, the coherent transport and manipulation of heat fluxes have been reported in Josephson junctions \cite{Giazotto12,Martinez15,Fornieri16} and quantum Hall systems \cite{Granger09,Altimiras10b,Altimiras12}. Intriguingly, the quantization of heat conductance has been observed in integer \cite{Jezouin13} and fractional quantum Hall systems \cite{Kane97,Banerjee16,Banerjee17_arxiv}, which were already known for the extremely precise quantization of their charge conductance. In this way, ample and valuable information about these peculiar states of matter, which was not accessible by charge measurement, is now available with interesting implications also for quantum computation \cite{Simon18,Wang18,Mross18}. New intriguing challenges posed by extending concepts like energy harvesting \cite{SanchezR15_PRL,SanchezR15_NJP,Samuelsson16,Thierschmann15,Juergens13,Erdman17,Mazza14}, driven heat and energy transport \cite{Ronetti17,Ludovico14,Ludovico16,Ludovico18,Mazza17}, energy exchange in open systems \cite{Carrega15,Carrega16} and fluctuation-dissipation theorems \cite{Campisi09,Campisi15,moskalets14,Averin10} to the quantum realm resulted in a great progress of the field of quantum thermodynamics.

A new perspective on EQO has been also triggered by the rising interest for heat transport properties of single-electron excitations. Mixed-charge correlators \cite{Crepieux14,Crepieux16,Battista14_jpcs} and heat fluctuations \cite{Battista13,Battista14} produced by single-electron sources were investigated and, in particular, it was shown that levitons are minimal excitations also for heat transport \cite{Vannucci17}. In addition, heat current has revealed a useful resource for the full reconstruction of a single-electron wave-function \cite{Moskalets16_pss}.

Here, we address the problem of the heat noise generated by levitons injected in a HOM interferometer in the fractional quantum Hall regime. We consider a four terminal quantum Hall bar in the Laughlin sequence \cite{Laughlin83}, where a single channel arises on each edge. Two terminals are contacted to time-dependent voltages, namely $V_L$ and $V_R$. Tunneling processes of quasi-particles are allowed by the presence of a QPC connecting the two edge states. In this case, charge noise generated in the HOM setup is identical to the one generated in a single-drive setup driven by the voltage $V_L-V_R$. Interestingly, we prove that this does not hold true anymore for heat noise, since it is possible to identify a contribution to HOM heat noise which is absent in a single-drive interferometer driven by $V_L-V_R$. In addition, we prove that the HOM heat noise always vanishes for a zero delay between the driving voltage, both for integer and fractional filling factors. Finally, we focus on the case of Lorentzian-shaped voltage carrying an integer number of electrons and we show that the HOM heat noise displays unexpected side dips in the fractional quantum Hall regime, which have no parallel in the integer regime. Intriguingly, the number of these side dips increases with the number of levitons injected per period. This result is consistent with the recently predicted phenomenon of charge crystallization of levitons in the fractional quantum Hall regime \cite{Ronetti18}.

The paper is organized as follows. In Sec. \ref{sec:model}, we introduce the model and the setup. Then, we evaluate charge and heat noises in Sec. \ref{sec:definitions}. In Sec. \ref{sec:results}, we present our results focusing on the peculiar case of levitons. Finally, we draw the conclusions in Sec. \ref{sec:conclusion}. Three Appendices are devoted to the technical aspects.

\section{Model \label{sec:model}}

\begin{figure}
	\centering
	\includegraphics[width=0.75\linewidth]{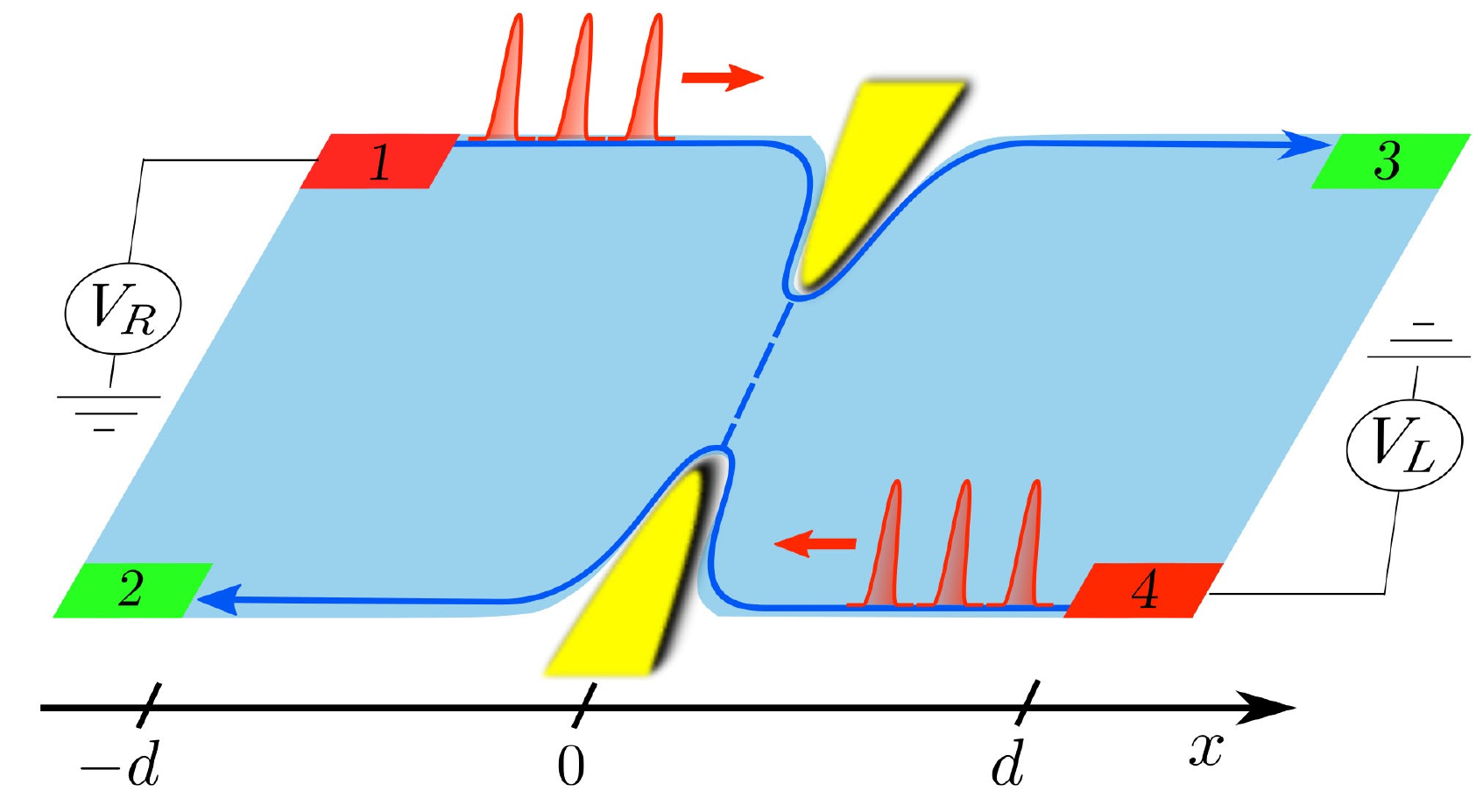}
	\caption{(Color online) Four-terminal setup for Hong-Ou-Mandel interferometry in the FQH regime. Contact 1 and 4 are used as input terminals, while contact 2 and 3 are the output terminals where current and noise are measured.}
	\label{fig:setup_hom}
\end{figure}

A quantum Hall bar in a four terminal geometry is depicted in Fig. \ref{fig:setup_hom}. In the Laughlin sequence $\nu=\frac{1}{2n+1}$, with integer $n \ge 0$, a single chiral mode arises on each edge \cite{Laughlin83,Wen90}. In the special case of integer quantum Hall effect at $\nu=1$ ($n=0$),  the system is composed by ordinary fermions and the chiral edge states are one-dimensional Fermi liquids. This description fails for other filling factors, where the excitations are quasi-particles with fractional charge $-\nu e$  (with $e>0$). The low-energy properties of the Laughlin states are well captured by an hydrodynamical model formulated in terms of right-moving and left-moving bosonic edge modes $\Phi_{R/L}(x)$, which satisfy commutation relations $\left[\Phi_{R/L}(x),\Phi_{R/L}(y)\right]=\pm i \pi \sign\left(x-y\right)$. The free Hamiltonian of these edge modes is (we set $\hbar=1$ throughout the paper) \cite{Wen95}
\begin{equation}
\label{free_Ham}
H_{0}=\frac{ v}{4\pi}\int dx \hspace{1mm}\sum\limits_{r=R,L}\left(\partial_x\Phi_{r}(x)\right)^2~,
\end{equation}
where $v$ is the velocity of propagation of right and left moving bosonic modes.\\ 
Terminals $1$ and $4$ are assumed to be connected to external time-dependent drives, while the remaining terminals are used to perform measurements. The charge densities, defined as
\begin{equation}
\label{eq:charge_dens}
\rho_{R/L}(x)=\pm \frac{e\sqrt{\nu}}{2\pi}\partial_x\Phi_{R/L}(x),
\end{equation} 
are capacitively coupled to the gate potentials $\mathcal{V}_{R/L}(x,t)$ through the following gate Hamiltonian \cite{Dolcini16,Dolcini18}
\begin{equation}
H_g=\int dx \hspace{1mm}\left\{\mathcal{V}_{R}(x,t)\rho_{R}(x)+\mathcal{V}_{L}(x,t)\rho_{L}(x)\right\}.
\end{equation}
The spatial dependence of the potentials is restricted to the region containing the semi-infinite contacts $1$ ($R$) and $4$ ($L$) by putting $\mathcal{V}_{R}(x,t)=\Theta(-(x+d))V_{R}(t)$ and $\mathcal{V}_{L}(x,t)=\Theta(x-d)V_{L}(t)$ (with $d>0$). Here, $V_{R/L}(t)=V_{R/L,dc}+V_{R/L,ac}(t)$ are periodic voltages, where $V_{R/L,dc}$ are time-independent dc components and $V_{R/L,ac}$ are pure periodic ac signals with period $\mathcal{T}=\frac{2\pi}{\omega}$, such that $\int_{0}^{\mathcal{T}}\frac{dt}{\mathcal{T}}V_{R/L}(t)=V_{R/L,dc}$. We remark that such modelization of the electromagnetic coupling between gate voltages and Hall bar occurs for gauge fixing with zero vector potential. 
\\
Since backscattering between the two edges is exponentially suppressed, we introduce a quantum point contact (QPC) at $x=0$, as shown in Fig. \ref{fig:setup_hom}, in order to allow for tunneling events between right- and left-moving excitations. We assume the QPC is tuned to a very low transparency, i.e. in the weak backscattering regime, where the tunneling of fractional quasi-particles is the only relevant process \cite{Kane92,Kane94,Safi10}. The corresponding additional term in the Hamiltonian is
\begin{equation}
\label{Ham:T}
H_{t}=\Lambda~ \Psi^{\dagger}_{R}(0)\Psi_{L}(0)+ \textit{H.c.}~,
\end{equation} 
where we introduced the quasi-particle fields represented by the bosonization identity \cite{giamarchi,Martin_Houches,Guyon02}
\begin{equation}
\label{eq:bosonization}
\Psi_{R/L}(x)=\frac{\mathcal{F}_{R/L}}{\sqrt{2\pi a}}e^{-i\sqrt{\nu}\Phi_{R/L}(x)},
\end{equation}
with $\mathcal{F}_{R/L}$ the so-called Klein factor, necessary for the proper anti-commutation relations, and $a$ the short-length cut-off.\\  
\section{Noises in the double-drive configuration \label{sec:definitions}}
The random partitioning, due to the poissonian tunneling at the QPC, generates fluctuations in the currents flowing along the quantum Hall bar. In this Section, we derive the expressions for charge and heat current noise in the double-drive configuration introduced in Sec. \ref{sec:model}, focusing on the regions downstream of the voltage contacts, namely $-d<x<d$.
\subsection{Charge noise}
We start by recalling the calculations for charge noise \cite{dubois13,Rech16,Glattli16_physE}. Charge current operators entering reservoirs $2$ and $3$ (located in $x=-d$ and $x=d$, respectively) can be expressed, due to chirality of Laughlin edge states, in terms of charge densities in Eq. \eqref{eq:charge_dens}
\begin{equation}
j_{2/3}(t)=\pm v \rho_{R/L}(\pm d,t). 
\end{equation}
The zero frequency cross-correlated charge noise is
\begin{equation}
	\mc S_{C} =  \int_{0}^{\mathcal{T}} \frac{dt}{\mathcal{T}} \int_{-\infty}^{+\infty} dt' \left[ \lan j_2(t') j_3(t) \ran - \lan j_2(t') \ran \lan j_3(t) \ran\right],
\end{equation}
where the thermal average is performed over the initial equilibrium density matrix, in absence of tunneling and driving voltage. In the weak backscattering regime, standard perturbative approach in the tunneling Hamiltonian will be used. The total time evolution of charge current operators with respect to $H_{0} + H_{g} + H_{t}$ can be then constructed in terms of powers of $\Lambda$ and reads
\begin{equation}
	 j_{2/3}(t) = j^{(0)}_{2/3}(t) + j^{(1)}_{2/3}(t) + j^{(2)}_{2/3}(t) + \mathcal{O}(\left|\Lambda\right|^3),
\end{equation}
with
\begin{align}
\label{eq:J0}
& \hspace{-3mm}	j^{(0)}_{2/3}(t) 
 = \pm v \rho_{R/L}^{(0)}(\pm d,t),\\
\label{eq:J1}
&\hspace{-3mm}	j^{(1)}_{2/3}(t)
 =\pm i v \int\limits_{-\infty}^t dt' \left[ H_{t}(t'),\rho_{R/L}^{(0)}(\pm d,t) \right],\\
\label{eq:J2}
&\hspace{-3mm}	j^{(2)}_{2/3}(t)
 = \pm (i)^2 v \int\limits_{-\infty}^t dt' \int\limits_{-\infty}^{t'} dt'' \left[  H_{t}(t'') , \left[ H_{t}(t'), \rho_{R/L}^{(0)}(\pm d,t)\right] \right],
\end{align}
where the tunneling Hamiltonian $H_{t}(t)$ and the charge densities $\rho_{R/L}^{(0)}(x,t)$ evolve in the interaction picture with respect to $H_0+H_g$. In order to make explicit the form of $\rho_{R/L}^{(0)}(x,t)$ it is sufficient to solve the equations of motion for the bosonic fields $\Phi_{R/L}$ with respect to $H_0+H_g$, i.e. in the absence of tunneling. The solutions read
\begin{equation}
\label{eq:sol_eq_motion}
	\Phi_{R/L}(x,t) = \phi_{R/L}(x,t) - e \sqrt \nu \int_{0}^{t\mp \frac x v -\frac d v } ds V_{R/L}(s),
\end{equation}
where $\phi_{R/L}(x,t)=\phi_{R/L}(x\mp vt)$ are the chiral bosonic fields at equilibrium (zero applied drive).\\
By exploiting the commutator
\begin{equation}
\left[H_t(t'),\rho^{(0)}_{R/L}(x,t)\right]=-\delta\left(t'-\left(t\mp\frac{x}{v}\right)\right)\dot{N}_{R/L}(x,t),
\end{equation}
where
\begin{align}
\dot{N}_{R}\left(x,t\right)&=i \nu e\Lambda  \Psi_{R}^{\dagger}(x-v t,0)\Psi_L(x-v t,0)+\text{H.c.},\\
\dot{N}_{L}\left(x,t\right)&=-i\nu e \Lambda  \Psi_{R}^{\dagger}(x+v t,0)\Psi_L(x+v t,0)+\text{H.c.},
\end{align} 
Eqs. \eqref{eq:J1} and \eqref{eq:J2} can be further recast as
\begin{align}
\label{j1b}
&j^{(1)}_{2/3}(t)=\dot{N}_{R/L}\left(\pm d,t\right),\\&\label{j2b}j^{(2)}_{2/3}(t)= i\int\limits_{-\infty}^{t-\frac{d}{v}}dt''\left[H_t(t''),\dot{N}_{R/L}\left(\pm d,t\right)\right].
\end{align}
In these expressions, we introduced the time evolution of quasi-particle fields with respect to $H_0+H_g$, which can be obtained from Eq. \eqref{eq:sol_eq_motion} using the bosonization identity 
\begin{align}
\label{quasi-particle_ev}
&\Psi_{R,L}(x,t)=\frac{\mathcal{F}_{R/L}}{\sqrt{2\pi a}}e^{-i\sqrt{\nu}\phi_{R/L}(x,t)}e^{i \nu e \int_{0}^{t\mp\frac{x}{v}-\frac{d}{v}}dt' V_{R/L}(t')},
\end{align} 
The current noise can be obtained from Eqs. \eqref{eq:J0}, \eqref{eq:J1}: the only non-vanishing contribution to second order in $\Lambda$ comes from $j^{(1)}_2(t+\tau) j^{(1)}_3(t)$, with terms $j^{(0)}_2(t+\tau) j^{(2)}_3(t)$ and $j^{(2)}_2(t+\tau) j^{(0)}_3(t)$ averaging to zero.\\
By introducing the correlator ($k_B=1$)
\begin{align}
P_{g}(t'-t)&= \langle e^{i\sqrt{g}\phi_{R/L}(0,t')}e^{-i\sqrt{g}\phi_{R/L}(0,t)}\rangle\nonumber=\\&=\left[ \frac{\pi  \theta \left(t'-t\right)}{\sinh\left( \pi  \theta  \left(t'-t\right) \right) \left(1+i \omega_c  \left(t'-t\right) \right)}\right]^{g},\label{eq:corr_func}
\end{align}
with $\theta$ the temperature and $\omega_c =v/a$ the high energy cut-off, one finds ($\lambda=\frac{\Lambda}{2\pi a}$)
\begin{align}
	&\mc S_{C}
	 =- 2 (\nu e)^2 |\lambda|^2 \int_{0}^{\mathcal{T}} \frac{dt}{\mathcal{T}} \int_{-\infty}^{+\infty} d t'  \times\nonumber\\&\times\cos \left\{ \nu e \int_{t}^{t'}  V_{-}(\tau) d \tau \right\} P_{2\nu}(t'-t),\label{eq:chargenoise}
\end{align}
where $V_{-}=V_R-V_L$.\\
Even though this charge noise is generated in a double-drive configuration, it is interesting to point out that it actually depends only on the single effective drive $V_{-}(t)$. The configuration with a single drive is usually termed in literature Hanbury-Brown-Twiss (HBT) setup \cite{HBT1,HBT2,Bocquillon12,dubois13-nature}. \\
Therefore, the charge noise presented in Eq. \eqref{eq:chargenoise} is the same as the one generated in a single-drive configuration, where reservoir 4 is grounded ($V_L(t)=0$) and reservoir 1 is contacted to the periodic voltage $V_{-}(t)$, such that
\begin{equation}
\label{rel:double-single}
S_{C}\left(V_R,V_L\right)=S_{C}\left(V_{-},0\right).
\end{equation}
Here, the arguments in brackets indicate the voltage applied to reservoirs $1$ and $4$, respectively. \\ One might consider Eq. \eqref{rel:double-single} as a consequence of a trivial shift of both voltages by a value corresponding to $V_L$. Nevertheless, such a result cannot be obtained by means of a gauge transformation (see Appendix \ref{gauge}). In this sense, Eq. \eqref{rel:double-single} implies that the charge noise \textit{incidentally} acquires the same expression in these two physically distinct experimental setups. As will be clearer in the following, for the charge case this is a consequence of the presence of a single local (energy independent) QPC. Generally, we expect that the double-drive and the single-drive ($V_R(t)=V_{-}(t)$ and $V_{L}(t)=0$) configurations return different outcomes for other physical observables, such as heat noise, as discussed in the next part.
\subsection{Heat noise \label{heatnoise}}
In the following, we evaluate the correlation noise of heat current between terminal $2$ and $3$ in the double-drive configuration. The heat current operators of terminal $2$ and $3$ can be expressed in terms of heat density operators \cite{kane96}
\begin{equation}
\mathcal{Q}_{R/L}(x,t)=\frac{v}{4\pi}\left(\partial_x\Phi_{R/L}(x,t)\right)^2,
\end{equation}
as
\begin{equation}
\mathcal{J}_{2/3}(t)=\pm v~ \mathcal{Q}_{R/L}(\pm d,t),
\end{equation}
due to the chirality of Laughlin edge states.\\ 
Then, we can define the cross-correlated heat noise
\begin{equation}
\mathcal{S}_{Q}=\int_{0}^{\mathcal{T}} \frac{dt}{\mathcal{T}}\int dt' \hspace{1mm}\left\{\langle \mathcal{J}_{2}(t')\mathcal{J}_{3}(t) \rangle-\langle \mathcal{J}_{2}(t')\rangle~\langle \mathcal{J}_{3}(t) \rangle\right\}~,
\end{equation}
Analogously to charge current, one can expand heat current operators in power of the tunneling amplitude $\Lambda$, thus obtaining
\begin{equation}
\label{curr_exp}
\mathcal{J}_{2/3}(t)=\mathcal{J}^{(0)}_{2/3}(t)+\mathcal{J}^{(1)}_{2/3}(t)+\mathcal{J}^{(2)}_{2/3}(t)+\mathcal{O}\left(\left|\Lambda\right|^3\right),
\end{equation}
where
\begin{align}
\label{J0}
&\mathcal{J}^{(0)}_{2/3}(t)=\pm v \mathcal{Q}^{(0)}_{R/L}(\pm d,t),\\&
\label{J1}
\mathcal{J}^{(1)}_{2/3}(t)=\pm iv \int\limits_{-\infty}^{t}dt'\left[H_t(t'),\mathcal{Q}^{(0)}_{R/L}(\pm d,t)\right],\\&\label{J2}\mathcal{J}^{(2)}_{2/3}(t)=\pm i^2 v\int\limits_{-\infty}^{t}dt'\int\limits_{-\infty}^{t'}dt''\left[H_t(t''),\left[H_t(t'),\mathcal{Q}^{(0)}_{R/L}(\pm d,t)\right]\right].
\end{align}
In the above equations we have denoted with $\mathcal{Q}^{(0)}(x,t)$, the time evolution of heat density in the absence of tunneling, which can be obtained from the time evolution of bosonic fields in Eq. \eqref{eq:sol_eq_motion} and reads 
\begin{align}
&\mathcal{Q}^{(0)}_{R/L}(x,t)=\frac{ v}{4\pi}\Big[\left(\partial_x\phi_{R/L}(x,t)\right)^2+\nonumber\\&\pm e\sqrt{\nu}\partial_x\phi_{R/L}(x,t)V_{R/L}\left(t\mp \frac{x}{v}\right)+ \frac{e^2\nu}{v}V^2_{R/L}\left(t\mp\frac{x}{v}\right)\Big]~.
\end{align}
The following commutator
\begin{equation}
\left[H_t(t'),\mathcal{Q}^{(0)}_{R/L}(x,t)\right]=-i\delta\left(t'-\left(t\mp\frac{x}{v}\right)\right)\dot{Q}_{R/L}(x,t),
\end{equation}
where
\begin{align}
\dot{Q}_{R}\left(x,t\right)&=v\Lambda \left(\partial_x \Psi^{\dagger}_{R}(x,t)\right)\Psi_L(x,t)+\text{H.c.},\\
\dot{Q}_{L}\left(x,t\right)&=-v\Lambda  \Psi^{\dagger}_{R}(x,t)\left(\partial_x\Psi_L(x,t)\right)+\text{H.c.},
\end{align} 
can be used to recast Eqs. \eqref{J1} and \eqref{J2}
\begin{align}
\label{J1b}
&\mathcal{J}^{(1)}_{2/3}(t)=\pm \dot{Q}_{R/L}\left(\pm d,t\right),\\&\label{J2b}\mathcal{J}^{(2)}_{2/3}(t)=\pm i\int\limits_{-\infty}^{t- \frac{d}{v}}dt''\left[H_t(t''),\dot{Q}_{R/L}\left(\pm d,t\right)\right].
\end{align}
The perturbative expansion of heat current operator in Eq. \eqref{curr_exp} allows to express heat correlation noise to lowest order as
\begin{equation}
\label{def:SQ}
\mathcal{S}_Q=\mathcal{S}^{(02)}_Q+\mathcal{S}^{(20)}_Q+\mathcal{S}^{(11)}_Q+\mathcal{O}\left(\left|\Lambda\right|^3\right),
\end{equation}
where
\begin{equation}
\label{eq:heatij}
\mathcal{S}_Q^{(ij)}=\int_{0}^{\mathcal{T}}\frac{dt}{\mathcal{T}}\int dt' \left\{ \langle \mathcal{J}_2^{(i)}(t')\mathcal{J}_3^{(j)}(t)\rangle-\langle \mathcal{J}_2^{(i)}(t')\rangle \langle\mathcal{J}_3^{(j)}(t)\rangle\right\}.
\end{equation}
Now, we can perform standard calculations, whose details are given in Appendix \ref{app:heat}, in order to evaluate all the terms appearing in Eq. \eqref{def:SQ}. By using the result of this calculation, it is possible to check whether an expression analogous to Eq. \eqref{rel:double-single} holds true also for heat noise. Interestingly, one finds that
\begin{equation}
\label{relHOMHBT}
\mathcal{S}_Q(V_R,V_L)=\mathcal{S}_Q(V_{-},0)+\Delta\mathcal{S}_Q(V_R,V_L),
\end{equation}
thus showing that, in contrast with the charge sector, heat fluctuations generated in the double-drive or in the single-drive configurations are different.
The two contributions in Eq. \eqref{relHOMHBT} are
\begin{widetext}
\begin{align}
\label{eq:SQHBT}
\mathcal{S}_Q(V_{-},0)&=|\lambda|^2\int_{0}^{\mathcal{T}}\frac{dt}{\mathcal{T}}\int dt'\Big\{\cos\left(\nu e\int_{t}^{t'}d\tau V_{-}(\tau)\right)\Re\left[\mathcal{P}_{\nu}(t'-t)\partial_{t}^2\mathcal{P}_{\nu}(t'-t)\right]\nonumber+\\&+\frac{\nu e v}{\pi}\int dt'' V_{-}(t') \mathcal{K}\left(t',t,t''\right)\sin\left(\nu e\int_{t}^{t'}d\tau V_{-}(\tau)\right)\Im\left[\partial_{t''}\mathcal{P}_{2\nu}(t''-t)\right]\Big\},\\\label{eq:SQ}
\Delta\mathcal{S}_Q(V_R,V_L)&=\nu^2 e^2|\lambda|^2\int_{0}^{\mathcal{T}}\frac{dt}{\mathcal{T}}\int dt'\cos\left(\nu e\int_{t}^{t'}d\tau V_{-}(\tau)\right)\Big(\alpha_{RL}(t,t')\Re\left[\mathcal{P}_{2\nu}(t'-t)\right]+\beta_{RL}(t,t')\Im\left[\mathcal{P}_{2\nu}(t'-t)\right]\Big),
\end{align}
\end{widetext}
where we defined the following functions
\begin{align}
&\mathcal{K}(t',t,t'')=\int d\tau \mathcal{P}_2(t'-\tau)\left(\Theta(\tau-t'')-\Theta(\tau-t)\right)=\nonumber\\&=\frac{\pi\theta}{v}\frac{\sinh\left(\pi \theta(t-t'')\right)}{\sinh\left(\pi \theta(t'-t)\right)\sinh\left(\pi \theta(t'-t'')\right)},\\
&\alpha_{RL}(t,t')=\left(V_R(t)V_L(t')-V_L(t)V_R(t')\right),\\&\beta_{RL}=\frac{v}{\pi}\int dt''\mathcal{K}(t'',t,t')V_{R}(t'')\left[V_L(t')-V_L(t)\right].\label{gamma}
\end{align}
The result of Eq. \eqref{relHOMHBT} arises because heat noise is sensitive to the energy distribution of the injected particles, thus leading to different outcomes in the single- and double-drive configurations. In this light, we expect this to hold true for general energy-dependent phenomena occurring at the QPC. For instance, any similarity between charge noises generated in the two setups discussed previously would disappear for more complicated tunneling geometry, such as multiple QPC or extended contacts, where transmission functions become energy-dependent \cite{Chevallier10,Dolcetto12,vannucci15,ronetti16}. \\ Eq. \eqref{relHOMHBT} further indicates that the double-drive and the single-drive configurations are completely distinct setups and that the relation in Eq. \eqref{rel:double-single} is solely a contingent effect of the single local QPC geometry.\\
It is useful to express heat correlation noise in energy space, by introducing the following Fourier series
\begin{align}
\label{ck}
\nu eV_{R/L}(t)&=\sum_{k} c_{k,R/L} e^{i k \omega t},\\
\label{pl}
e^{-i \nu e\int_{0}^{t}d\tau V_{-}(\tau)}&=\sum_{l} \tilde{p}_l e^{-i (l+q_R-q_L) \omega t},
\end{align}
where we defined also the number of particles excited by $V_{R/L}$ along the system in a period
\begin{equation}
q_{R/L}=\frac{\nu e}{2\pi}\int_{0}^{\mathcal{T}}dt\hspace{1mm} V_{R/L}(t)=\frac{\nu e V_{R/L,dc}}{\omega},
\end{equation}
and the Fourier transform of $P_g(t)$ in Eq. \eqref{eq:corr_func}
\begin{align}
\tilde{\mathcal{P}}_{g}(E)&=\int dt \mathcal{P}_g(t)e^{i E t}=\nonumber\\&=\left(\frac{2 \pi \theta}{\omega_c}\right)^{g-1}\frac{e^{\frac{E}{2\theta}}}{\Gamma(g)\omega_c}\left|\Gamma\left(\frac{g}{2}-i \frac{E}{2\pi \theta}\right)\right|^2.
\end{align}
By exploiting these results, the two contributions to $\mathcal{S}_Q$ become
\begin{widetext}
\begin{align}
&\mathcal{S}_Q(V_{-},0)=-\left|\lambda\right|^2\sum_{l}\Big\{\frac{2\nu^2 \pi^2 \theta^2+(1+\nu)\left((l+q_R-q_L)\omega\right)^2}{1+2\nu}\left|\tilde{p}_l\right|^2\tilde{\mathcal{P}}_{2\nu}((l+q_R-q_L)\omega)+\nonumber\\&-\frac{1}{4}\sum_{k\ne 0}\left(c_{k,R}-c_{k,L}\right)\left(\tilde{p}_{l-k}\tilde{p}^{*}_{l}-\tilde{p}_{l}\tilde{p}^{*}_{l+k}\right)(l+q_R-q_L)\omega \coth\frac{k\omega}{2\theta}\left(\tilde{\mathcal{P}}_{2\nu}((l+q_R-q_L)\omega)-\tilde{\mathcal{P}}_{2\nu}(-(l+q_R-q_L)\omega)\right)\Big\},\label{eq:SHBTenergy}\\
&\Delta\mathcal{S}_Q(V_R,V_L)=\frac{\left|\lambda\right|^2}{2}\sum_{k,p,l}\left(c_{k,R}c_{p,L}-c_{k,L}c_{p,R}\right)\tilde{p}_{l+k+p}\tilde{p}^{*}_{l}\frac{\mathcal{W}_{(l+q_R-q_L),k,p}+\mathcal{W}_{(l+q_R-q_L),p,k}}{2},\label{eq:SQenergy}
\end{align}
where the coefficients $\mathcal{W}_{l,k,p}$ encodes all the effects due to temperature and interaction on $\Delta\mathcal{S}_Q$ and reads
\begin{align}
\mathcal{W}_{l,k,p}=\frac{\omega_c}{4\pi}\int\frac{dE}{2\pi}\Big\{\tilde{\mathcal{P}}_{1}(E)&\tilde{\mathcal{P}}_{1}(k\omega-E)\Big[\tilde{\mathcal{P}}_{2\nu-1}(E-l\omega)+\tilde{\mathcal{P}}_{2\nu-1}(-E-(l+k+p)\omega)+\tilde{\mathcal{P}}_{2\nu-1}(-E+l\omega)\Big]+\nonumber\\&+\tilde{\mathcal{P}}_{2\nu-1}(E+(l+k+p)\omega)\Big\}-\frac{1}{2}\left(\tilde{\mathcal{P}}_{2\nu}((l+k)\omega)+\tilde{\mathcal{P}}_{2\nu}(-(l+k)\right).
\end{align}
\end{widetext}
Let us observe that the contribution $\Delta\mathcal{S}_Q$ exists only in the double-drive configurations. Indeed, in the configuration with a single drive, where $V_L=0$, one obtains that $c_{k,L}=0$ for each $k$, and the contribution in Eq. \eqref{eq:SQenergy} vanishes.
\subsection{Hong-Ou-Mandel noises}
Among all the possible choices for the configuration involving the two voltages $V_R$ and $V_L$, one of the most interesting, even from the experimental point of view, is the Hong-Ou-Mandel (HOM) setup, where two identical voltage drives are applied to reservoirs $1$ and $4$ and delayed by a constant time $t_D$. This experimental configuration corresponds to set $V_R(t)=V(t)$ and $V_L(t)=V(t+t_D)$ in Eq. \eqref{eq:chargenoise}, with $V(t)$ a generic periodic drive. In this situation the charge excited by each drive along the edge channels are equal, such that $q_R=q_L=q$.\\ 
For notational convenience, we define the single-drive heat noise and the HOM charge and heat noises as 
\begin{align}
\label{def:HBT}
\mathcal{S}^{sd}_{Q}&=\mathcal{S}_{Q}(V_{-}(t),0),\\
\mathcal{S}^{HOM}_{C/Q}&=\mathcal{S}_{C/Q}(V(t),V(t+t_D)).\label{def:HOM}
\end{align}
According to Eq. \eqref{relHOMHBT} and using the above definitions, the HOM heat noise can be expressed as
\begin{equation}
\label{rel:HOM-HBT}
\mathcal{S}_{Q}^{HOM}=\mathcal{S}_{Q}^{sd}+\Delta\mathcal{S}_{Q}.
\end{equation}
From the existing literature \cite{dubois13,Rech16,Ronetti18}, it is well established that charge HOM noise reduces to its equilibrium value for null time delay. Before entering into the details of our discussion, we would like to prove analytically that the same holds true for HOM heat noise $\mathcal{S}_Q^{HOM}$, independently of the choice of any parameter. The photo-assisted amplitude in Eq. \eqref{pl} reduces to $\tilde{p}_l=\delta_{l,0}$ and the Fourier coefficients $c_{k,-}$ vanish for all $k$. Let us start by looking at the single-drive contribution. By substituting this analytical simplification in Eq. \eqref{eq:SHBTenergy}, we obtain
\begin{equation}
\label{HBTzerotd}
\mathcal{S}_Q^{sd}(t_D=0)=-\left|\lambda\right|^2\frac{\nu^2 \pi^2 \theta^2}{1+2\nu}\equiv \mathcal{S}_Q^{vac},
\end{equation} 
which is independent of the injected particles and correspond simply to the equilibrium noise $\mathcal{S}_Q^{vac}$ due to thermal fluctuations. This can be clearly understood given the fact that $V_{-}(t)=0$ for $t_D=0$ and the single-drive contribution corresponds to the noise generated in a driveless configuration. \\
Concerning the remaining part in Eq. \eqref{rel:HOM-HBT}, one has for $t_D=0$
\begin{align}
\label{SQtD}
\Delta\mathcal{S}_Q=\frac{\left|\lambda\right|^2}{4}\sum_{k}c_{k,R}c_{-k,R}\left(\mathcal{W}_{0,k,-k}+\mathcal{W}_{0,-k,k}\right),
\end{align}
where 
\begin{align}
\label{wkk}
\mathcal{W}_{0,k,-k}=\frac{\tilde{\mathcal{P}}_{2\nu}(k\omega)-\tilde{\mathcal{P}}_{2\nu}(-k\omega)}{2}.
\end{align}
From Eq. \eqref{wkk}, we can clearly deduce that $\mathcal{W}_{0,k,-k}=-\mathcal{W}_{0,-k,k}$, which enforces the vanishing of $\Delta\mathcal{S}_Q$ in Eq. \eqref{SQtD}. This is enough to prove that HOM heat noise always reaches its equilibrium value at $t_D=0$, such that
\begin{equation}
\label{HOMzerotd}
\mathcal{S}_Q^{HOM}(t_D=0)=\mathcal{S}_Q^{sd}(t_D=0)= \mathcal{S}_Q^{vac}.
\end{equation} 
Let us note that this is not a trivial result since $\mathcal{S}_Q^{HOM}$ does not depend effectively on the single drive $V_{-}$ as $\mathcal{S}_Q^{sd}$, but on both $V_R$ and $V_L$ and even at $t_D=0$ the system is still driven by these two voltages.
\section{Results and discussions \label{sec:results}}
In this section, we discuss the results concerning the heat correlation noises in the HOM interferometer. In particular, we focus our discussion on a specific driving voltage, namely a periodic train of Lorentzian pulses
\begin{equation}
V_{Lor}(t)=\frac{V_0}{\pi}\sum\limits_{k=-\infty}^{+\infty}\frac{W}{W^2+(t-k\mathcal{T})^2}.
\end{equation}
A Lorentzian-shaped drive, which satisfy the additional quantization condition
\begin{equation}
\label{lor_quant}
\nu e\int_{0}^{\mathcal{T}}dt\hspace{1mm}V_{Lor}(t)=2\pi q,
\end{equation}
with $q$ an integer number, constitutes the optimal driving able to inject clean pulses devoid of any additional electron-hole pairs. The minimal excitations thus emitted into the quantum Hall channels are the aforementioned levitons \cite{levitov96,keeling06}. The Fourier coefficients for this specific drive are given in Appendix \ref{app:fourier}.\\ 
In the HOM setup previously described, a state composed by $q_L=q_R=q$ levitons \cite{Moskalets18} is injected by each driven contact and collide at the QPC, separated by a controllable time delay.\\ 
In analogy with the previous literature on charge noise, we introduce the following ratio \cite{Wahl14,Ferraro13,Rech16,Marguerite16}
\begin{equation}
\label{ratioHOM}
\mathcal{R}_{C/Q}^{HOM}=\frac{\mathcal{S}^{HOM}_{C/Q}-\mathcal{S}_{C/Q}^{vac}}{2\mathcal{S}^{R}_{C/Q}-2\mathcal{S}_{C/Q}^{vac}},
\end{equation}
where we subtracted the equilibrium noise $\mathcal{S}_{C/Q}^{vac}$ and we normalize with respect to $\mathcal{S}_{C/Q}^{R}\equiv \mathcal{S}_{C/Q}(V_R,0)$, which are charge and heat noises expected for the random partitioning of a single source of levitons, i.e. when $V_R(t)=V_{Lor}(t)$ and $V_L(t)=0$. The expressions for $\mathcal{S}_{C}^{vac}$ and $\mathcal{S}_{C}^{R}$ are well-known and have been derived in previous paper \cite{Rech16,Vannucci17,dubois13}. The expression for $\mathcal{S}_Q^{R}$ can be obtained from our results in Sec. \ref{heatnoise} and reads
\begin{align}
&\hspace{-3mm}\mathcal{S}_Q^{R}=-\left|\lambda\right|^2\sum_{l}\Big\{\frac{2\nu^2 \pi^2 \theta^2+(1+\nu)\left(l\omega\right)^2}{1+2\nu}\left|p_l\right|^2\tilde{\mathcal{P}}_{2\nu}((l+q)\omega)\nonumber+\\&\hspace{-5mm}-\sum_{k\ne 0}c_{k}\left(p_{l-k}p^{*}_{l}-p_{l}p^{*}_{l+k}\right)(l+q)\omega \frac{\tilde{\mathcal{P}}_{2}(k\omega)}{2 k \omega}\times\nonumber\\&\times\left(\tilde{\mathcal{P}}_{2\nu}((l+q)\omega)-\tilde{\mathcal{P}}_{2\nu}(-(l+q)\omega)\right)\Big\},
\end{align}
where $c_{k}=\nu e\int_{0}^{\mathcal{T}}\frac{dt}{\mathcal{T}}V_{Lor}(t)e^{i k \omega t}$ are the Fourier coefficients for a single Lorentzian voltage and $p_l=\int_{0}^{\mathcal{T}}\frac{dt}{\mathcal{T}} e^{-i \nu e\int_{0}^{t}d\tau V_{lor}(\tau)} e^{i(l+q)\omega t}$ (see Appendix \ref{app:fourier}).\\
In addition, we define an analogous ratio for single-drive heat noise as
\begin{equation}
\label{ratioHBT}
\mathcal{R}_Q^{sd}=\frac{\mathcal{S}^{sd}_Q-\mathcal{S}_Q^{vac}}{2\mathcal{S}^{R}_{Q}-2\mathcal{S}_Q^{vac}},
\end{equation}
in order to asses its relative contribution to the overall HOM heat noise.\\ Let us notice that, according to Eqs \eqref{HBTzerotd} and \eqref{HOMzerotd} both ratios, $\mathcal{R}_Q^{HOM}$ and $\mathcal{R}_Q^{sd}$ vanish for $t_D=0$. In the specific case of levitons, which are single-electron excitations, at $\nu=1$ the physical explanation for the total dip at $t_D=0$ involves the anti-bunching effect of identical fermions: electron-like excitations colliding at the QPC at the same time are forced to escape on opposite channels, thus leading to a total suppression of fluctuations at $t_D=0$ and generating the so called Pauli dip \cite{dubois13-nature,Jonckheere12,Bocquillon12}. For fractional filling factors, it is remarkable that this total dip is still present despite the presence of anyonic quasi-particles in the system, which do not obey Fermi-Pauli statistics \cite{Rech16,Ferraro18}. Anyway, this single QPC geometry does not allow for the braiding of one quasi-particle around the other, thus excluding any possible effect due to fractional statistics.\\
 \begin{figure}[h]
	\centering
	\includegraphics[width=1\linewidth]{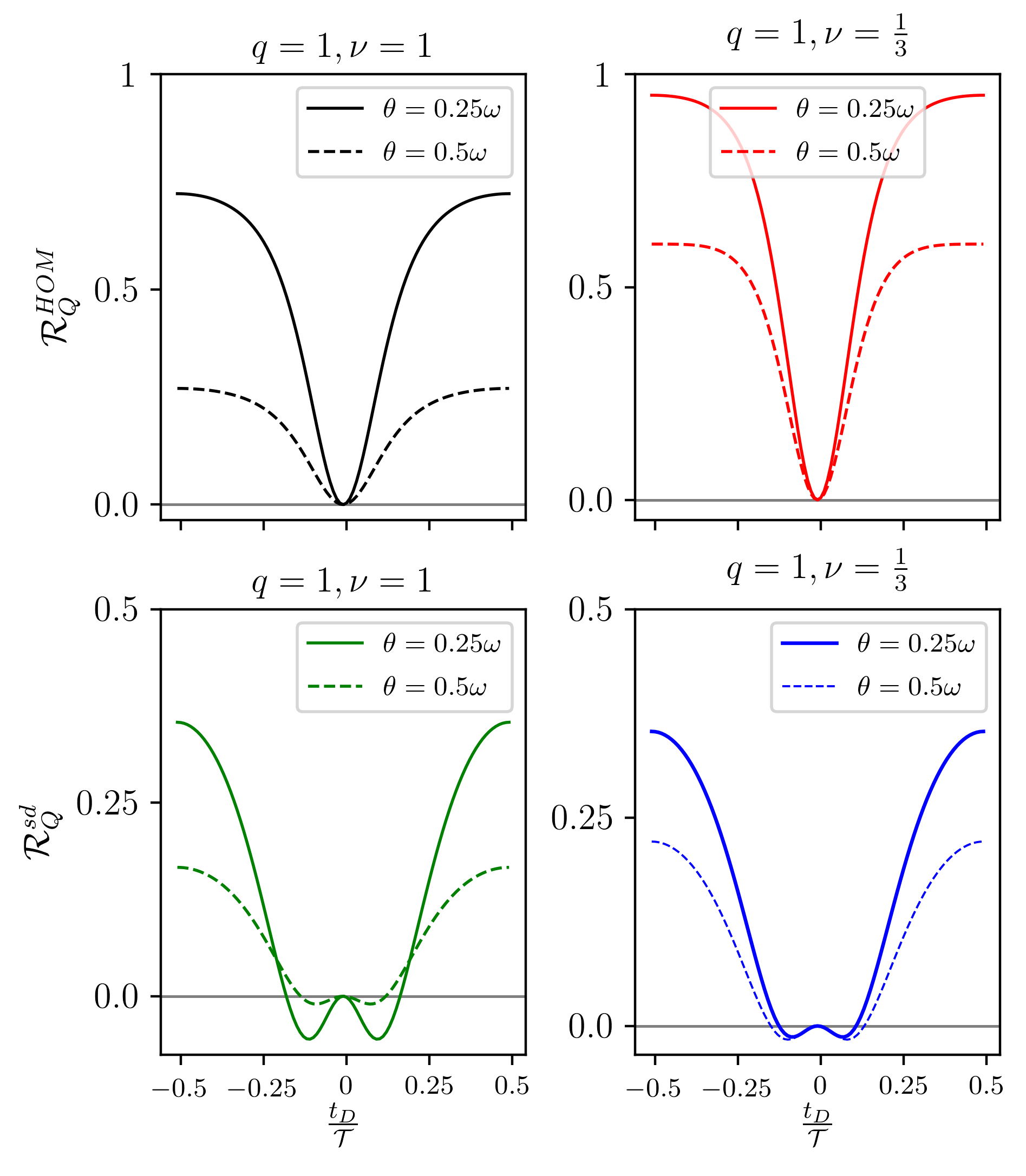}
	\caption{(Color online) HOM heat ratio $\mathcal{R}_Q^{HOM}$ (upper panels) and single-drive heat ratio $\mathcal{R}_Q^{sd}$ (lower panels) as a function of the time delay $t_D$ for $q=1$ and temperatures $\theta=0.25\omega$ (solid lines) and $\theta=0.5\omega$ (dashed lines). The integer case (left panel) and the fractional case for $\nu=\frac{1}{3}$ (right panel) are compared. The other parameters are $W=0.1 \mathcal{T}$ and $\omega=0.01 \omega_c$.}
	\label{fig:ratio1}
\end{figure}In the following, we exploit the full generality of our derivation by performing the analysis for different values of $q$.\\ 
We start by considering the regime where thermal and quantum fluctuations are comparable.\\ As a beginning, we focus on the relevant case of $q=1$, where states formed by a single leviton are injected from both sources \cite{Moskalets16_pss}. The collision of identical single-leviton states is very interesting because previous work on fluctuations of charge current proved that in this case the ratio of HOM charge noise is independent of filling factors and temperatures, acquiring an universal analytical expression \cite{Rech16,Glattli16_physE}. In order to perform a similar comparison for the heat noise, we present in Fig. \ref{fig:ratio1} the HOM heat ratio (upper panels) considering two temperatures $\theta=0.25\omega$ (solid line) and $\theta=0.5\omega$ (dashed lines) for both the integer and fractional case. Contrarily to the charge case, these curves are all clearly distinct. This means that this universality does not extend also to heat fluctuations. This fact can be explained by the dependence of heat HOM noise on the energy distribution of particles injected by the drives, which in turn is significantly affected by the temperature and by the strength of correlations encoded in the filling factor $\nu$. In particular, as the temperature is further increased, the thermal fluctuations tend to hide the effect of the voltages, resulting in a reduction of $\mathcal{R}_{Q}^{HOM}$ for both filling factors.\\ Interestingly, we also note that the single-drive ratio can switch sign as $t_D$ is tuned, independently of the filling factor. Since $\mathcal{S}_{Q}^{R}$ is independent of $t_D$, the change of sign of $\mathcal{R}_{Q}^{sd}$ is entirely due to $\mathcal{S}^{sd}_Q$ itself. This is a remarkable difference with respect to the charge noise generated in the same configurations, since charge conservation fixes the sign of current-current correlations. On the contrary, it should be pointed out that the sign of heat noise is not constrained by any conservation law \cite{moskalets14}.\\
\begin{figure}[h]
	\centering
	\includegraphics[width=1\linewidth]{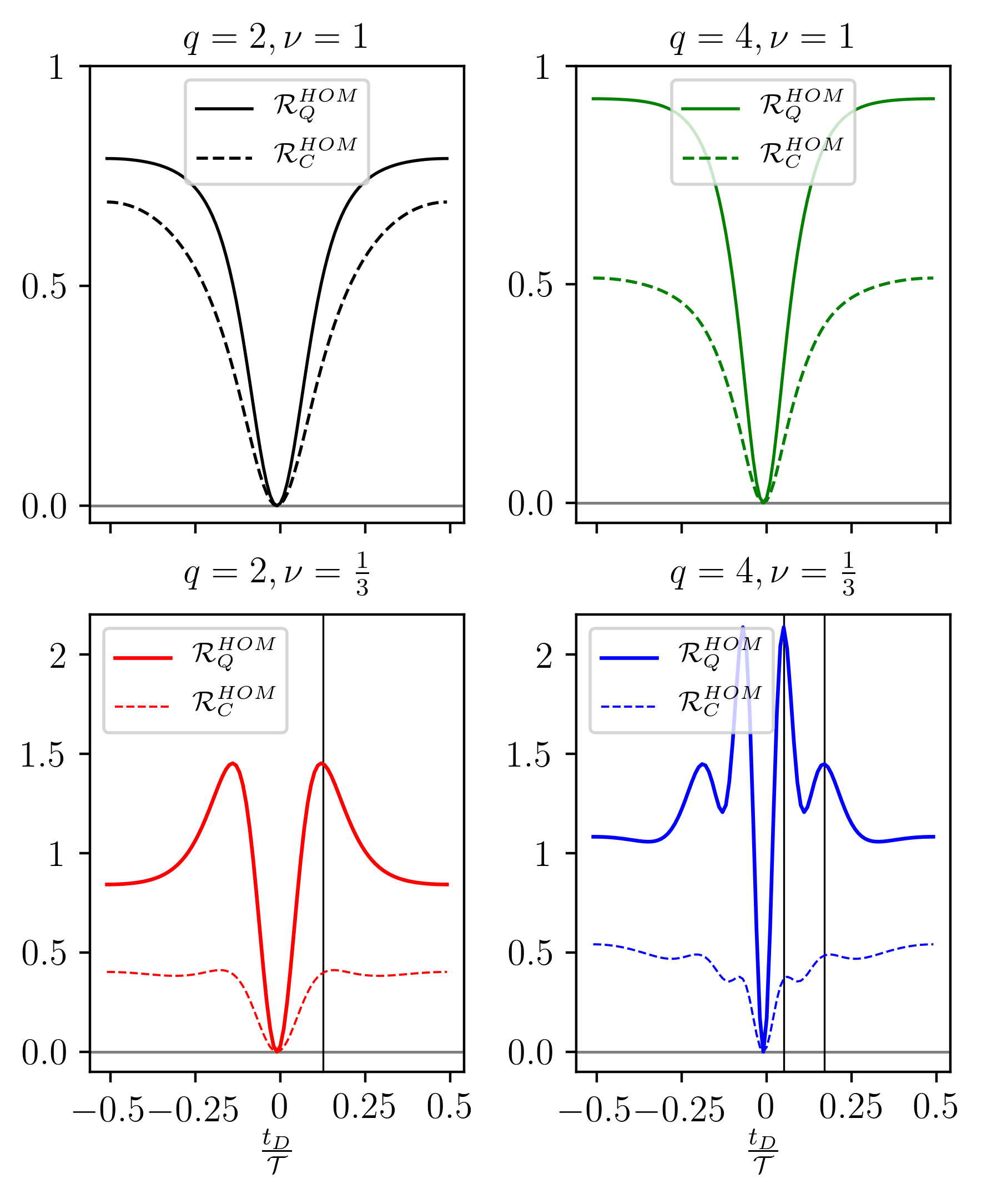}
	\caption{HOM heat ratio $\mathcal{R}_Q^{HOM}$ (solid lines) and HOM charge ratio $\mathcal{R}_C^{HOM}$ (dashed lines) as a function of the time delay $t_D$ for $q=2$ and $q=4$. The integer case (upper panels) and the fractional case for $\nu=\frac{1}{3}$ (lower panels) are compared. Black vertical lines demonstrate the exact correspondence of side peaks appearing in charge and heat ratio. The other parameters are $W=0.1 \mathcal{T}$, $\theta=0.25 \omega$ and $\omega=0.01 \omega_c$.}
	\label{fig:ratio2}
\end{figure}In Fig. \ref{fig:ratio2}, we start looking at the collision of states composed by multiple levitons and compare HOM charge and heat ratios (solid and dashed lines, respectively) for $q=2$ and $q=4$. In the fermionic case, presented in the two upper panels, both charge and heat ratio show a single smooth dip at $t_D=0$, without additional side features. Interestingly, heat fluctuations are enhanced with respect to charge: in particular, heat HOM ratios saturate to their asymptotic value for smaller values of time delay compared to charge ratio. Again, the enhancement of heat fluctuations can be related to the fact that heat is not constrained by any conservation law, in contrast to the case of charge.\\ Very remarkably, the curves for the HOM ratio in the fractional case display instead some unexpected side peaks and dips in addition to the central dip. In particular, the number of these maxima and minima increases for states composed with more levitons. A recent paper by the authors explained this intriguing result for charge HOM noise in terms of a crystallization process induced by strong correlation on the charge density of $q$ levitons , i.e. a re-arrangement of the density into an oscillating and ordered pattern with a number of peaks related to $q$ \cite{Ronetti18}. Black vertical lines in the lower panel of Fig. \ref{fig:ratio2} demonstrate the exact correspondence of side peaks appearing in charge and heat ratio as a function of time delay. Based on this argument, we can infer that the HOM heat noise is affected by the crystallization induced in the propagating levitons, thus giving rise to the features observed in the lower panel of Fig. \ref{fig:ratio2}. While the oscillating pattern of $\mathcal{R}_Q^{HOM}$ remarkably matches with that of $\mathcal{R}_C^{HOM}$, the amplitude oscillations are widely enhanced for heat fluctuations, in particular for the peaks occurring at small values of time delay.\\
\begin{figure}
	\centering
	\includegraphics[width=1\linewidth]{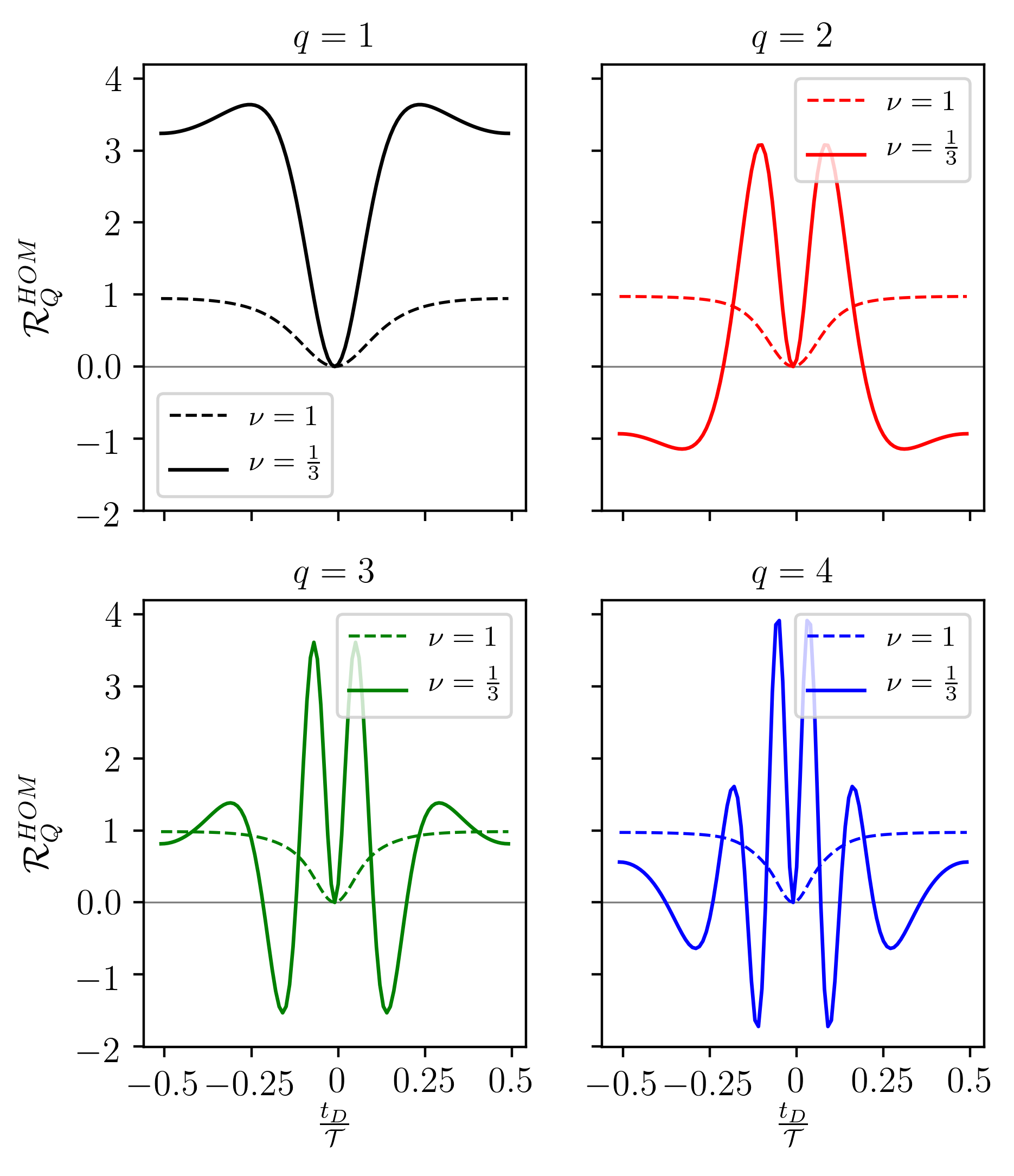}
	\caption{(Color online) HOM heat ratio $\mathcal{R}_Q^{HOM}$ as a function of the time delay $t_D$ for $q=1$, $q=2$, $q=3$, $q=4$. The integer case (dashed lines) and the fractional case for $\nu=\frac{1}{3}$ (solid lines) are compared. The other parameters are $W=0.1 \mathcal{T}$, $\theta=10^{-4}\omega$ and $\omega=0.01 \omega_c$.}
	\label{fig:ratio3}
\end{figure}
We conclude by noticing that strong correlation of the fractional regime can increase the value of the HOM heat ratio even above $1$. Once again, since this is not the case for the single-drive contribution, this is due to the presence of $\Delta\mathcal{S}_Q$, which is peculiar to collision between levitons incoming from different reservoirs.\\ 
Now, we consider the regime of very low temperature $\theta\ll\omega$, where the quantum effects should be largely enhanced with respect to the thermal fluctuations. Having established from the previous discussion the connection between $\Delta\mathcal{S}_Q$ and $\mathcal{S}_Q^{HOM}$ in the fractional regime, we focus only on the HOM heat ratio $\mathcal{R}_Q^{HOM}$.\\ The plots for $\mathcal{R}_{Q}^{HOM}$ in the integer and in the fractional case are compared in Fig. \ref{fig:ratio3} for different values of $q$. In the integer case, a single smooth dip is present for all the values of $q$, confirming the phenomenology described for the finite temperature case. For the strongly correlated case, at $q=1$ one observes a smooth profile, except for a small decrease close to $t_D=0.5$. Intriguingly, the oscillations observed in Fig. \ref{fig:ratio2} for $q>1$ are widely enhanced in this regime, such that the HOM ratio displays zeros, whose number increases with $q$, in addition to the central one and can also reach negative values.
\section{Conclusion \label{sec:conclusion}}

In this work, we investigated charge and heat current fluctuations in an HOM interferometer in the fractional quantum Hall regime. Here, two identical leviton excitations impinge at a QPC with a given time delay. We started by evaluating zero-frequency cross-correlated charge and heat noises in the presence of two generic driving voltage $V_L$ and $V_R$. We demonstrated that heat noise in this double-drive configuration depends on both $V_+=V_L+V_R$ and $V_-=V_L-V_R$ and, thus, cannot be reproduced in a single-drive setup driven by the voltage $V_-$ only. In particular, this implies that single-drive configuration and HOM interferometer implemented with voltage sources are two physically distinct experimental configurations. Moreover, we proved that the HOM heat ratio vanishes for a null time delay for both integer and fractional filling factors, despite the presence, in the latter case, of emergent fractionally charged quasi-particles. Finally, we investigated the form of HOM heat ratio for different regimes of temperatures. Interestingly, unexpected side dips emerged only in the fractional regime which can be related to the crystallization mechanism recently predicted for levitons \cite{Ronetti18}.

\begin{acknowledgments}
L.V.\ and M.S. acknowledge support from CNR SPIN through Seed project ``Electron quantum optics with quantized energy packets''. This work was granted access to the HPC resources of Aix-Marseille Universit\'e financed by the project Equip@Meso (Grant No. ANR-10-EQPX-29-01). It has been carried out in the framework of project ``1shot reloaded'' (Grant No. ANR-14-CE32-0017) and benefited from the support of the Labex ARCHIMEDE (Grant No. ANR-11-LABX-0033), all funded by the ``investissements d'avenir'' French Government program managed by the French National Research Agency (ANR). The project leading to this publication has received funding from Excellence Initiative of Aix-Marseille University - A*MIDEX, a French ``investissements d'avenir'' programme.
\end{acknowledgments}
\appendix

\section{Coupling to the gate \label{gauge}}
In this Appendix, we show that there is no gauge transformation able to link the equations of motion for the configurations with two driving voltages $V_R$ and $V_L$ and the configuration with a single drive $V_{-}=V_R-V_L$, presented in the main text.\\
In the double-drive setup a voltage drive is applied both to right-moving and left-moving excitations. We consider a situation in which the vector potentials $\mathcal{A}_{R/L}(x,t)$ are absent. The Lagrangian density is
\begin{widetext}
\begin{align}	
	\mc L
	& = \frac{1}{4\pi} \bigg\{ - \partial_x\Phi_R(x,t) \big[\partial_t \Phi_R(x,t) + v \partial_x\Phi_R(x,t)\big] + \partial_x\Phi_L(x,t) \big[\partial_t \Phi_L(x,t) - v \partial_x\Phi_L(x,t)\big] \bigg\} + \nonumber \\
	& \quad + \frac{e \sqrt{\nu}}{2\pi} \bigg[ \partial_x \Phi_R(x,t) \mc V_R(x,t) - \partial_x \Phi_L(x,t) \mc V_L(x,t) \bigg].
\end{align}
\end{widetext}
The Euler-Lagrange equations
\begin{equation}
\label{eq:Euler-Lagrange}
	\partial_t \frac{\delta \mc L}{\delta \partial_t \Phi_\alpha} + \partial_x \frac{\delta \mc L}{\delta \partial_x \Phi_\alpha} - \frac{\delta \mc L}{\delta \Phi_\alpha} = 0 
\end{equation}
with $\alpha = R,L$, give rise to the following equation of motions for the bosonic fields:
\begin{align}
	& (\partial_t + v \partial_x) \Phi_R(x,t) = e \sqrt \nu \mc V_R(x,t) \label{app:eq_motion1} \\
	& (\partial_t - v \partial_x) \Phi_L(x,t) = e \sqrt \nu \mc V_L(x,t). \label{app:eq_motion2}
\end{align}
In order to model the system presented in Sec. \ref{sec:model}, the form for the voltage drives is
\begin{align}
	\mc V_R(x,t) & = f_R(x)V_R(t) \\
	\mc V_L(x,t) & = f_L(x)V_L(t)
\end{align}
where $f_{R/L}(x)$ are time-independent, while $V_{R/L}(t)$ are space-independent. 
In this case equation of motions for the double-drive setup are
\begin{subequations}
\label{eq:eq_motion_HOM}
\begin{align}
	& (\partial_t + v \partial_x) \Phi_R(x,t) = e \sqrt \nu f_R(x)V_R(t) \\
	& (\partial_t - v \partial_x) \Phi_L(x,t) = e \sqrt \nu f_L(x)V_L(t).
\end{align}
\end{subequations}
We also consider a single-drive setup with an effective voltage drive $\mc V_R(x,t)=f_R(x)[V_R(t)-V_L(t)]$ on the right side, and the left side grounded [$\mc V_L(x,t)=0$]. We still consider that the magnetic potential is zero on both edges. It is immediate to show that the equation of motions are now
\begin{subequations}
\label{eq:eq_motion_HBT}
\begin{align}
	& (\partial_t + v \partial_x) \Phi_R(x,t) = e \sqrt \nu f_R(x)[V_R(t)-V_L(t)] \\
	& (\partial_t - v \partial_x) \Phi_L(x,t) = 0
\end{align}
\end{subequations}

\subsection{Applying gauge transformations to the HOM setup}
Here we show that a gauge transformation that operates in the following way on the voltage drives
\begin{equation}
	\begin{cases}
		\mc V_R(x,t) = f_R(x)V_R(t) \\
		\mc V_L(x,t) = f_L(x)V_L(t)
	\end{cases} \longrightarrow 
	\begin{cases}
		\mc V'_R(x,t) = f_R(x)[V_R(t)-V_L(t)]\\
		\mc V'_L(x,t) = 0,
	\end{cases}
\end{equation}
does not transform Eqs. \eqref{eq:eq_motion_HOM} into Eqs. \eqref{eq:eq_motion_HBT}, but leaves them unchanged.\\
We recall that a general gauge transformation that leaves invariant an electromagnetic field is given by
\begin{align}
	\mc V'_{R/L}(x,t) & = \mc V_{R/L}(x,t) - \partial_t \chi_R(x,t), \\
	\mc A'_{R/L}(x,t) & = \mc A_{R/L}(x,t) + \partial_x \chi_R(x,t),
\end{align}
with $\chi_{R/L}(x,t)$ a scalar function.\\
In our particular case, voltage potentials are required to transform as
\begin{align}
	\mc V'_R(x,t) & = f_R(x) V_R(x) - \partial_t \chi_R(x,t) = f_R(x)[V_R(t)-V_L(t)] \\
	\mc V'_L(x,t) & = f_L(x) V_L(x) - \partial_t \chi_L(x,t) = 0
\end{align}
for the right-moving and left-moving sector respectively. The transformation is evidently implemented by the choice
\begin{subequations}
\label{eq:HOM->HBT}
\begin{align}
	\chi_R(x,t) & = f_R(x) \int_{0}^t d \tau V_L(\tau) \\
	\chi_L(x,t) & = f_L(x) \int_{0}^t d \tau V_L(\tau)
\end{align}
\end{subequations}
Since these equations involve spatial-dependent functions, we expect that non-zero magnetic potentials arise as a consequence of the gauge transformation. In the new gauge we get non-zero magnetic potentials given by (in our initial gauge choice $\mc A_{R/L}=0$)
\begin{align}
	\mc A'_R(x,t) & = \partial_x f_R(x) \int_{0}^t d \tau V_L(\tau) \\
	\mc A'_L(x,t) & = \partial_x f_L(x) \int_{0}^t d \tau V_L(\tau)
\end{align}
and the Lagrangian density now reads
\begin{widetext}
\begin{align}	
	\mc L'
	& = \frac{1}{4\pi} \bigg\{ - \partial_x\Phi_R(x,t) \big[\partial_t \Phi_R(x,t) + v \partial_x\Phi_R(x,t)\big] + \partial_x\Phi_L(x,t) \big[\partial_t \Phi_L(x,t) - v \partial_x\Phi_L(x,t)\big] \bigg\} + \nonumber \\
	& \quad + \frac{e \sqrt{\nu}}{2\pi}  \bigg\{ \partial_x \Phi_R(x,t) f_R(x)[V_R(t)-V_L(t)] + \big[ \partial_t \Phi_R(x,t) \partial_x f_R(x) - \partial_t \Phi_L(x,t) \partial_x f_L(x) \big]  \int_{t_0}^t d \tau V_L(\tau) \bigg\}
\end{align}
\end{widetext}
where the last term accounts for the presence of $\mc A'_R(x,t)$ and $\mc A'_L(x,t)$. We now look for the equation of motions in this new configuration. From Euler-Lagrange equations one gets
\begin{align}
	& (\partial_t + v \partial_x) \Phi_R(x,t) =  \nonumber\\&=e \sqrt \nu f_R(x)[V_R(t)-V_L(t)]+ e \sqrt \nu f_R(x) V_L(t)=\nonumber\\& = e \sqrt \nu f_R(x) V_R(t) \\
	& (\partial_t - v \partial_x) \Phi_L(x,t) = e \sqrt \nu f_L(x)V_L(t).
\end{align}
Note that we have not recovered the equation of motions for the single drive setup, Eqs.\ \eqref{eq:eq_motion_HBT}, as one may naively expect. On the contrary, we have found the equations of motion for the double-drive setup, Eqs.\ \eqref{eq:eq_motion_HOM}. 
\section{Heat noise \label{app:heat}}
In this Appendix, we give more details about the calculation of heat noise presented in Sec. \ref{sec:definitions}. Before starting with the derivation of heat noise, we would give some formulas that would be useful in the following parts.
\subsection{Useful formulas \label{useful}}
In the following, we derive some results that would be useful for the evaluation of heat current fluctuations. In particular, our goal is to evaluate the following average values (for simplicity, we drop all the low indices $R$ or $L$)
\begin{align}
&C_1(t_1,t_2,t_3)=\langle \partial_{t_1}\phi(t_1)e^{i\sqrt{ \nu}\phi(t_2)}e^{-i\sqrt{ \nu}\phi(t_3)}\rangle,\label{ExpA}\\
&C_2(t_1,t_2,t_3)=\langle e^{i\sqrt{ \nu}\phi(t_1)}e^{-i\sqrt{ \nu}\phi(t_2)}\partial_{t_3}\phi(t_3)\rangle,\label{ExpA'}\\
&D_1(t_1,t_2,t_3)=\langle \left(\partial_{t_1}\phi(t_1)\right)^2e^{i\sqrt{ \nu}\phi(t_2)}e^{-i\sqrt{ \nu}\phi(t_3)}\rangle,\label{ExpB}\\
&D_2(t_1,t_2,t_3)=\langle e^{i\sqrt{ \nu}\phi(t_1)}e^{-i\sqrt{ \nu}\phi(t_2)}\left(\partial_{t_3}\phi(t_3)\right)^2\rangle,\label{ExpB'}
\end{align}
where the thermal average is performed over the initial equilibrium density matrix, in absence of tunneling and driving voltage and bosonic fields evolve according to the edge Hamiltonian $H_0$.
In order to evaluate $C_1$ and $C_2$, we start by considering the following general average value
\begin{equation}
\label{E1}
E_1(\epsilon_1,\epsilon_2,\epsilon_3;t_1,t_2,t_3)=\langle e^{-i\epsilon_1\phi(t_1)}e^{-i\epsilon_2\phi(t_2)}e^{-i\epsilon_3\phi(t_3)}\rangle,
\end{equation}
which is connected to $C_1$ and $C_2$ by this relation
\begin{align}
\label{connection1}
C_1(t_1,t_2,t_3)=i\partial_{t_1}\left\{\lim\limits_{\epsilon_1\rightarrow0}\partial_{\epsilon_1}E_1(\epsilon_1,\epsilon_2,\epsilon_3;t_1,t_2,t_3)\right\}_{\substack{\epsilon_2=-\sqrt{ \nu}\\\epsilon_3=\sqrt{ \nu}}},\\
\label{connection1'}
C_2(t_1,t_2,t_3)=i\partial_{t_3}\left\{\lim\limits_{\epsilon_3\rightarrow0}\partial_{\epsilon_3}E_1(\epsilon_1,\epsilon_2,\epsilon_3;t_1,t_2,t_3)\right\}_{\substack{\epsilon_1=-\sqrt{ \nu}\\\epsilon_2=\sqrt{ \nu}}}.
\end{align}
By using \cite{vondelft}
\begin{align}
\label{VD}
\langle e^{ \chi(t_1)}e^{\chi(t_2)}e^{\chi(t_3)}\rangle=e^{\frac{1}{2}\sum_{i=1}^{3}\langle \chi(t_i)^2\rangle} e^{\sum_{i<j}\langle \chi(t_i) \chi(t_j)\rangle},
\end{align}
we obtain from Eq. \eqref{E1}
\begin{align}
&E_1(\epsilon_1,\epsilon_2,\epsilon_3;x_1,x_2,x_3)=e^{-\frac{1}{2}\sum_{i=1}^{3}\langle \epsilon_i^2\phi^2(x_i)\rangle} \times\nonumber\\&\times e^{-\left\{\epsilon_1\epsilon_2\langle\phi(x_1)\phi(x_2)\rangle+\epsilon_1\epsilon_3\langle\phi(x_1)\phi(x_3)\rangle+\epsilon_2\epsilon_3\langle\phi(x_2)\phi(x_3)\rangle \right\}}.
\end{align}
Finally, we use Eqs. \eqref{connection1} and \eqref{connection1'} to find $C_1$ and $C_2$
\begin{align}
\label{relC1}
C_1(t_1,t_2,t_3)&=-i\sqrt{ \nu}\mathcal{K}(t_1,t_2,t_3)P_{\nu}(t_2-t_3),\\
\label{relC2}
C_2(t_1,t_2,t_3)&=-i\sqrt{ \nu}\mathcal{K}(-t_3,-t_1,-t_2)P_{\nu}(t_1-t_2),
\end{align}
where we defined (see Eq. \eqref{eq:corr_func} in the main text)
\begin{align}
\label{Pdt}
P_{g}(t'-t)&= \langle e^{i\sqrt{g}\phi_{R/L}(0,t')}e^{-i\sqrt{g}\phi_{R/L}(0,t)}\rangle\nonumber=\\&=\left[ \frac{\pi  \theta \tau}{\sinh\left( \pi  \theta \tau \right) \left(1+i \omega_c \tau \right)}\right]^{g},
\end{align}
and
\begin{align}
\mathcal{K}(t_1,t_2,t_3)&=\partial_{t_1}\left\{\langle \phi(t_1)\phi(t_3)\rangle-\langle \phi(t_1)\phi(t_2)\rangle\right\}\nonumber=\\&=\int d\tau \mathcal{P}_2(t_1-\tau)\left(\Theta(\tau-t_3)-\Theta(\tau-t_2)\right).\label{K}
\end{align}
One could also obtain the following similar relations
\begin{align}
\label{relC1b}
&\langle \partial_{t_1}\phi(t_1)e^{-i\sqrt{ \nu}\phi(t_2)}e^{i\sqrt{ \nu}\phi(t_3)}\rangle=i\sqrt{ \nu}\mathcal{K}(t_1,t_2,t_3)P_{\nu}(t_2-t_3),\\
\label{relC2b}
&\langle e^{-i\sqrt{ \nu}\phi(t_1)}e^{i\sqrt{ \nu}\phi(t_2)\partial_{t_3}\phi(t_3)}\rangle=i\sqrt{ \nu}\mathcal{K}(-t_3,-t_1,-t_2)P_{\nu}(t_1-t_2).
\end{align}
Exploiting the following average
\begin{equation}
\langle\partial_{t_1} \phi(t_1)\partial_{t}\phi(t)\rangle=-\frac{\pi^2 \theta^2}{v^2\sinh^2\left(\pi \theta (t_1-t)\right)}
\end{equation}
the function $\mathcal{K}$ can be further evaluated by using
\begin{align}
\partial_{t_1}\langle \phi(t_1)\phi(t_2)\rangle&=\int_{-\infty}^{t_2}dt\langle\partial_{t_1} \phi(t_1)\partial_{t}\phi(t)\rangle\nonumber=\\&=\frac{\pi \theta}{v}\left[\coth\left(\pi \theta(t_1-t_2)\right)-1\right].
\end{align}
By using this result, one finds
\begin{align}
\mathcal{K}(t_1,t_2,t_3&)=\frac{\pi \theta}{v}\left(\coth\left(\pi \theta(t_1-t_3)\right)-\coth\left(\pi \theta(t_1-t_2)\right)\right)=\nonumber\\=&
\frac{\pi \theta}{v}\frac{\sinh\left(\pi \theta(t_2-t_3)\right)}{\sinh\left(\pi \theta(t_1-t_3)\right)\sinh\left(\pi \theta(t_1-t_2)\right)}.\label{finalK}
\end{align}
In order to evaluate $D_1$ and $D_2$, we start by considering the following general average value
\begin{align}
\label{E2}
&E_2(\epsilon_1,\epsilon_2,\epsilon_3,\epsilon_4;t_1,t_2,t_3,t_4)=\nonumber\\&=\langle e^{-i\epsilon_1\phi(t_1)}e^{-i\epsilon_2\phi(t_2)}e^{-i\epsilon_3\phi(t_3)}e^{-i\epsilon_4\phi(t_4)}\rangle,
\end{align}
which is connected to $D_1$ and $D_2$ by these relations
\begin{widetext}
\begin{align}
\label{connection2}
D_1(t_1,t_2,t_3)&=-\partial_{t_1}\partial_{t'_1}\left\{\lim\limits_{\epsilon_1\rightarrow0, \epsilon_2\rightarrow 0}\partial_{\epsilon_1}\partial_{\epsilon_2}E_2(\epsilon_1,\epsilon_2,\epsilon_3,\epsilon_4;t_1,t'_1,t_2,t_3)\right\}_{\substack{\epsilon_1=-\epsilon_2=-\sqrt{ \nu}\\\hspace{-8.5mm}t'_1=t_1}},\\
\label{connection2'}
D_2(t_1,t_2,t_3)&=-\partial_{t_3}\partial_{t'_3}\left\{\lim\limits_{\epsilon_1\rightarrow0, \epsilon_2\rightarrow 0}\partial_{\epsilon_3}\partial_{\epsilon_4}E_2(\epsilon_1,\epsilon_2,\epsilon_3,\epsilon_4;t_1,t_2,t_3,t'_3)\right\}_{\substack{\epsilon_4=-\epsilon_3=\sqrt{ \nu}\\\hspace{-8.5mm}t'_3=t_3}},\\
\end{align}
\end{widetext}
\begin{widetext}
By using Eq. \eqref{VD}, we obtain from Eq. \eqref{E1}
\begin{align}
&E_2(\epsilon_1,\epsilon_2,\epsilon_3,\epsilon_4;t_1,t_2,t_3,t_4)=e^{-\frac{1}{2}\sum_{i=1}^{4} \epsilon_i^2\langle\phi^2(t_i)\rangle}\cdot\\&\cdot e^{-\left\{\epsilon_1\epsilon_2\langle \phi(t_1)\phi(t_2)\rangle+\epsilon_1\epsilon_3\langle\phi(t_1)\phi(t_3)\rangle+\epsilon_1\epsilon_4\langle\phi(t_1)\phi(t_4)\rangle+\epsilon_2\epsilon_3\langle\phi(t_2)\phi(t_3)\rangle+\epsilon_2\epsilon_4\langle\phi(t_2)\phi(t_4)\rangle+\epsilon_3\epsilon_4\langle\phi(t_3)\phi(t_4)\rangle \right\}}.
\end{align}
Finally, we use Eq. \eqref{connection2} and \eqref{connection2'} to find $D_1$ and $D_2$
\begin{align}
\label{relB}
\left(\partial_{t_1}\phi(t_1)\right)^2e^{i\sqrt{ \nu}\phi(t_2)}e^{-i\sqrt{ \nu}\phi(t_3)}\rangle&=\left\{\langle \left(\partial_{t_1}\phi(t_1)\right)^2\rangle-\nu\left(\mathcal{K}(t_1,t_2,t_3)\right)^2\right\} P_{\nu}(t_2-t_3),\\\label{relB'}
\hspace{2mm}\langle e^{i\sqrt{ \nu}\phi(t_1)}e^{-i\sqrt{ \nu}\phi(t_2)}\left(\partial_{t_3}\phi(t_3)\right)^2\rangle&=\left\{\langle \left(\partial_{t_3}\phi(t_3)\right)^2\rangle-\nu\left(\mathcal{K}(-t_3,-t_1,-t_2)\right)^2\right\} P_{\nu}(t_1-t_2).
\end{align}
By carrying on a similar calculation, one can find also the analogous quantities
\begin{align}
\label{relB2}
\left(\partial_{t_1}\phi(t_1)\right)^2e^{-i\sqrt{ \nu}\phi(t_2)}e^{i\sqrt{ \nu}\phi(t_3)}\rangle&=\left\{\langle \left(\partial_{t_1}\phi(t_1)\right)^2\rangle-\nu\left(\mathcal{K}(t_1,t_2,t_3)\right)^2\right\} P_{\nu}(t_2-t_3),\\\label{relB2'}
\hspace{2mm}\langle e^{-i\sqrt{ \nu}\phi(t_1)}e^{i\sqrt{ \nu}\phi(t_2)}\left(\partial_{t_3}\phi(t_3)\right)^2\rangle&=\left\{\langle \left(\partial_{t_3}\phi(t_3)\right)^2\rangle-\nu\left(\mathcal{K}(-t_3,-t_1,-t_2)\right)^2\right\} P_{\nu}(t_1-t_2).
\end{align}
\end{widetext}
\subsection{Calculations of heat noise}
Our starting point is the perturbative expression of heat noise given in the main text (see Eq. \eqref{curr_exp})
\begin{equation}
\mathcal{S}_Q=\mathcal{S}_Q^{(02)}+\mathcal{S}_Q^{(20)}+\mathcal{S}_Q^{(11)}+\mathcal{O}\left(\left|\Lambda\right|^3\right).
\end{equation}
Firstly, we derive the term $\mathcal{S}_Q^{(11)}$, which reads
\begin{align}
\mathcal{S}_Q^{(11)}&=\int_{0}^{\mathcal{T}}\frac{dt}{\mathcal{T}}\int dt' \Big\{ \langle \partial_{t'}\Psi^{\dagger}_R(0,t')\Psi_L(0,t')\partial_t\Psi^{\dagger}_L(0,t)\Psi_R(0,t)\rangle\nonumber+\\&+\langle \Psi^{\dagger}_L(0,t')\partial_{t'}\Psi_R(0,t')\Psi^{\dagger}_R(0,t)\partial_t\Psi_L(0,t)\rangle\Big\},
\end{align}
since $\langle\mathcal{J}_{2/3}^{(1)}(t)\rangle=0$ (see Eq. \eqref{J1b} in the main text). We recall  that the time evolution of quasi-particle fields is
\begin{equation}
\Psi_{R,L}(x,t)=\frac{\mathcal{F}_{R/L}}{\sqrt{2\pi a}}e^{-i\sqrt{\nu}\phi_{R/L}(x,t)} e^{i \nu e \int_{t_0}^{t\mp\frac{x}{v}-\frac{d}{v}}dt' V_{R/L}(t')}.
\end{equation} 
We can further express the average in the above equation as
\begin{widetext}
\begin{align}
\label{Sigma11_2}
&\mathcal{S}_Q^{(11)}= 2\left|\lambda\right|^2\int_{0}^{\mathcal{T}} \frac{dt}{\mathcal{T}}\int dt' \hspace{1mm}\Big\{\cos{\left(\nu e\int_{t'}^{t}dt''V_R(t'')-V_L(t'')\right)}\partial_t'\mathcal{P}_{\nu}(t'-t)\partial_t\mathcal{P}_{\nu}(t'-t)+\\&+\nu e V_R(t')\sin{\left(\nu e\int_{t'}^{t}dt''V_R(t'')-V_L(t'')\right)}\frac{1}{2}\partial_t\mathcal{P}_{2\nu}(t'-t)+\nu e V_L(t)\sin{\left(\nu e\int_{t'}^{t}dt''V_R(t'')-V_L(t'')\right)}\frac{1}{2}\partial_{t'}\mathcal{P}_{2\nu}(t'-t)+\nonumber\\&-\nu^2 e^2 V_R(t')V_L(t)\cos{\left(\nu e\int_{t'}^{t}dt''V_R(t'')-V_L(t'')\right)}\mathcal{P}_{2\nu}(t'-t)\Big\}~,\nonumber
\end{align}
where the function $P_{g}(t)$ is defined in Eq. \eqref{Pdt} and $\lambda\equiv\frac{\Lambda}{2\pi a}$. The integration by parts of second and third line of Eq. \eqref{Sigma11_2} provides some useful eliminations, providing the final expression for this contribution
\begin{align}
\label{Sigma11_3}
&\mathcal{S}_Q^{(11)}
=2\left|\lambda\right|^2\int_{0}^{\mathcal{T}} \frac{dt}{\mathcal{T}}\int dt' \hspace{1mm}\Big\{\cos{\left(\nu e\int_{t'}^{t}dt''\left(V_R(t'')-V_L(t'')\right)\right)}\partial_t'\mathcal{P}_{\nu}(t'-t)\partial_t\mathcal{P}_{\nu}(t'-t)+\\&-\frac{1}{2}\nu^2 e^2 \left(V_R(t')V_R(t)+V_L(t')V_L(t)\right)\cos{\left(\nu e\int_{t'}^{t}dt''\left(V_R(t'')-V_L(t'')\right)\right)}\mathcal{P}_{2\nu}(t'-t)\Big\}~.
\end{align}
\end{widetext}
We focus on the remaining contributions, starting from $\mathcal{S}_Q^{(02)}$: the calculations for the other term would be analogous. By plugging Eqs. \eqref{J0} and \eqref{J2} in the definition of $\mathcal{S}_Q^{(02)}$, one finds
\begin{widetext}
\begin{align}
\mathcal{S}_Q^{(02)}&=-i\frac{\left|\lambda\right|^2}{4\pi}\int dt\int_{0}^{\mathcal{T}}\frac{dt'}{\mathcal{T}}\int dt''\theta(t'-t'')\Big\{\langle\left(\partial_{t}\phi_R(0,t)\right)^2\left[\Psi^{\dagger}_{R}(0,t'')\Psi_L(0,t''),\partial_{t'}\Psi^{\dagger}_L(0,t')\Psi_R(0,t')\right]\rangle\nonumber+\\&-2\nu e V_R(t)\langle\partial_{t}\phi_R(0,t)\left[\Psi^{\dagger}_{R}(0,t'')\Psi_L(0,t''),\partial_{t'}\Psi^{\dagger}_L(0,t')\Psi_R(0,t')\right]\rangle+\nonumber\\&-\langle\left(\partial_{t}\phi_R(0,t)\right)^2\rangle\langle\left[\Psi^{\dagger}_{R}(0,t'')\Psi_L(0,t''),\partial_{t'}\Psi^{\dagger}_L(0,t')\Psi_R(0,t')\right]\rangle\Big\}.
\end{align}
\end{widetext}
The averages involving the commutators can be performed by using the expression in Eq. \eqref{quasi-particle_ev} for the time evolution of quasi-particle fields and by resorting to the formulas in Eqs \eqref{relC1}, \eqref{relC1b},\eqref{relB} and \eqref{relB'} derived in the Appendix \ref{useful}. Indeed, one finds
\begin{widetext}
\begin{align}
&\mathcal{S}^{(02)}_{(Q)}=-i\frac{\left|\lambda\right|^2}{4\pi}\int dt\int_{0}^{\mathcal{T}} \frac{dt'}{\mathcal{T}}\int dt''~\Theta\left(t'-t''\right)\Big\{-\Big[\nu\partial_{t'}\mathcal{K}^2(t,t',t'')\cos{\left(\nu e\int_{t''}^{t'}d\tau V_-(\tau)\right)}\left(\mathcal{P}_{2\nu}(t''-t')-\mathcal{P}_{2\nu}(t'-t'')\right) \nonumber+\\
&+\nu eV_L(t')\mathcal{K}^2(t,t',t'')\left(\mathcal{P}_{2\nu}(t''-t')-\mathcal{P}_{2\nu}(t'-t'')\right)\sin\left({\nu e\int_{t''}^{t'}d\tau V_-(\tau)}\right)\Big]+\nonumber\\&+\mathcal{K}(t,t',t'')\Big[2\nu eV_R(t)\sin\left({\nu e\int_{t''}^{t'}d\tau V_-(\tau)}\right)\partial_{t'}\left[\mathcal{P}_{2\nu}\left(t''-t'\right)-\mathcal{P}_{2\nu}\left(t'-t''\right)\right]+\nonumber\\
& -4\nu^2 e^2V_R(t)V_L(t')\cos\left({\nu e\int_{t''}^{t'}d\tau V_-(\tau)}\right)\left[\mathcal{P}_{2\nu}\left(t''-t'\right)-\mathcal{P}_{2\nu}\left(t'-t''\right)\right]\Big]\Big\}\nonumber,
\end{align}
A similar calculation leads to the expression for the last contribution, given by
\begin{align}
&\mathcal{S}^{(20)}_{(Q)}=-i\frac{\left|\lambda\right|^2}{4\pi}\int dt\int_{0}^{\mathcal{T}} \frac{dt'}{\mathcal{T}}\int dt''~\Theta\left(t'-t''\right)\Big\{\Big[\nu\partial_{t'}\mathcal{K}^2(t,t',t'')\cos{\left(\nu e\int_{t''}^{t'}d\tau V_-(\tau)\right)}\left(\mathcal{P}_{2\nu}(t''-t')-\mathcal{P}_{2\nu}(t'-t'')\right) \nonumber+\\
&+\nu eV_L(t')\mathcal{K}^2(t,t',t'')\left(\mathcal{P}_{2\nu}(t''-t')-\mathcal{P}_{2\nu}(t'-t'')\right)\sin\left({\nu e\int_{t''}^{t'}d\tau V_-(\tau)}\right)\Big]+\nonumber\\&+\mathcal{K}(t,t'',t')\Big[2\nu eV_L(t)\sin\left({\nu e\int_{t''}^{t'}d\tau V_-(\tau)}\right)\partial_{t'}\left[\mathcal{P}_{2\nu}\left(t''-t'\right)-\mathcal{P}_{2\nu}\left(t'-t''\right)\right]+\nonumber\\
& +4\nu^2 e^2V_R(t)V_L(t'')\cos\left({\nu e\int_{t''}^{t'}d\tau V_-(\tau)}\right)\left[\mathcal{P}_{2\nu}\left(t''-t'\right)-\mathcal{P}_{2\nu}\left(t'-t''\right)\right]\Big]\Big\}\nonumber.
\end{align}
By summing up the two contributions, one can see that the first lines cancel out and the remaining two lines add up in a way that allows to get rid of the function $\Theta(t'-t'')$, thus obtaining
\begin{align}
&\mathcal{S}^{(02)}_{(Q)}+\mathcal{S}^{(20)}_{(Q)}=-i\frac{\left|\lambda\right|^2}{4\pi}\int dt\int_{0}^{\mathcal{T}} \frac{dt'}{\mathcal{T}}\int dt''~\Big\{\mathcal{K}(t,t'',t')\Big[2\nu eV_-(t)\sin\left({\nu e\int_{t''}^{t'}d\tau V_-(\tau)}\right)\partial_{t'}\left[\mathcal{P}_{2\nu}\left(t''-t'\right)-\mathcal{P}_{2\nu}\left(t'-t''\right)\right]+\nonumber\\
&-4\nu^2 e^2V_R(t)V_L(t'')\cos\left({\nu e\int_{t''}^{t'}d\tau V_-(\tau)}\right)\left[\mathcal{P}_{2\nu}\left(t''-t'\right)-\mathcal{P}_{2\nu}\left(t'-t''\right)\right]\Big]\Big\}\nonumber.
\end{align}
\end{widetext}
Now, summing all the contributions according to Eq. \eqref{relHOMHBT}, it is possible to obtain the result presented in the main text, which reads
\begin{equation}
\mathcal{S}_Q(V_R,V_L)=\mathcal{S}_Q(V_{-},0)+\Delta\mathcal{S}_Q(V_R,V_L),
\end{equation}
with
\begin{widetext}
	\begin{align}
	\label{eq:SQHBT_app}
	\mathcal{S}_Q(V_{-},0)&=|\lambda|^2\int_{0}^{\mathcal{T}}\frac{dt}{\mathcal{T}}\int dt'\Big\{\cos\left(\nu e\int_{t}^{t'}d\tau V_{-}(\tau)\right)\Re\left[\mathcal{P}_{\nu}(t'-t)\partial_{t}^2\mathcal{P}_{\nu}(t'-t)\right]\nonumber+\\&+\frac{\nu e v}{\pi}\int dt'' V_{-}(t') \mathcal{K}\left(t',t,t''\right)\sin\left(\nu e\int_{t}^{t'}d\tau V_{-}(\tau)\right)\Im\left[\partial_{t''}\mathcal{P}_{2\nu}(t''-t)\right]\Big\},\\\label{eq:SQ_app}
	\Delta\mathcal{S}_Q(V_R,V_L)&=\nu^2 e^2|\lambda|^2\int_{0}^{\mathcal{T}}\frac{dt}{\mathcal{T}}\int dt'\cos\left(\nu e\int_{t}^{t'}d\tau V_{-}(\tau)\right)\Big(\alpha_{RL}(t,t')\Re\left[\mathcal{P}_{2\nu}(t'-t)\right]+\beta_{RL}(t,t')\Im\left[\mathcal{P}_{2\nu}(t'-t)\right]\Big),
	\end{align}
\end{widetext}
where we defined the following functions
\begin{align}
&\alpha_{RL}(t,t')=\left(V_R(t)V_L(t')-V_L(t)V_R(t')\right),\\&\beta_{RL}=\frac{v}{\pi}\int dt''\mathcal{K}(t'',t,t')V_{R}(t'')\left[V_L(t')-V_L(t)\right].\label{gamma_app}
\end{align}

\section{Fourier coefficients}
\label{app:fourier}
This Appendix is devoted to the Fourier analysis of the Lorentzian periodic signal $V_{Lor}(t)$ and of the phase $e^{-i\nu e\int_{t_0}^{t}dt' V_{Lor}(t')}$, where
\begin{align}
		V_{Lor}(t) & = \frac{V_{0}}{\pi} \sum_{k=-\infty}^{+\infty} \frac{W}{W^2 + (t-k\mathcal{T})^2} ,
\end{align}
where $\mathcal{T}$ is the periodic, $V_0$ the amplitude and $W$ the half width at half maximum.\\ The coefficients for the Fourier series of the expression $\nu eV_{Lor}(t)=\sum_{k}c_{k}e^{i k \omega t}$ are
\begin{equation}
\label{eq_app:lorcoeff}
c_{k}=q \omega e^{-2\pi \frac{W}{\mathcal{T}}|k|},
\end{equation}
with $q=\frac{\nu e}{2\pi}\int_{0}^{\mathcal{T}}dt  V_{Lor}(t)=\frac{\nu e V_0}{\omega}$.\\
We also note that, for the time delayed voltage $V_{Lor}(t+t_D)$, the coefficients become $c'_{k}=c_{k}e^{-ik \omega t_D}$.\\
The Fourier series $e^{-i\nu e\int_{0}^{t}dt' (V_{Lor}(t')-V_0)} = \sum_l p_l e^{-i l\omega t}$ allows to deal with the time-dependent problem as a superposition of time-independent configurations, with energy shifted by an integer amount of energy quanta $\omega$.
For the Lorentzian case, it is convenient to switch to a complex representation in terms of the variable $z=e^{i \omega t}$. After some algebra and introducing $\gamma=e^{-2\pi \eta}$ one finds \cite{dubois13,grenier13}
\begin{equation}
	\label{eq:pl_lorentz}
	p_l = \frac{1}{2\pi i} \oint_{|z|=1} dz \, z^{l+q-1} \left( \frac{1-z\gamma}{z- \gamma} \right)^{q} .
\end{equation}
From Eq.~\eqref{eq:pl_lorentz} one can make use of complex binomial series and Cauchy's integral theorem \cite{arfken,needham} to finally get
\begin{equation}
	p_l = q \gamma^l \sum_{s=0}^\infty (-1)^{s} \frac{\Gamma(l+s+q)}{\Gamma(1+q-s)} \frac{\gamma^{2s}}{s! (s+l)!} .
\end{equation}
Finally, the Fourier coefficients $\tilde{p}_l$ for the voltage phase $e^{-i \nu e\int_{0}^{t}d\tau \left(V_{Lor}(\tau)-V_{Lor}(\tau+t_D)\right)}$ in the HOM configuration are given by
\begin{equation}
\tilde{p}_l=\int_{0}^{\mathcal{T}}\frac{dt}{\mathcal{T}}e^{i l \omega t}e^{-i \nu e\int_{0}^{t}d\tau \left(V_{Lor}(\tau)-V_{Lor}(\tau+t_D)\right)},
\end{equation}
which can be calculated in terms of the coefficient $p_l$ of a single drive as
\begin{equation}
\tilde{p}_l=\sum_{m}p^{*}_{m}p_{m+l} e^{-i m \omega t_D}.
\end{equation}


%

%

\end{document}